\documentclass[aps,preprint,showpacs]{revtex4}

\usepackage[pdftex]{graphicx}
\usepackage{latexsym}
\usepackage{amsbsy}
\usepackage{amsmath}
\usepackage{amssymb}

\begin{document}

\date{\today}

\title{Self-Consistent Field Theory studies of the thermodynamics and
  quantum spin dynamics of magnetic Skyrmions}    

\author{R.\ Wieser}
\affiliation{1. International Center for Quantum Materials, Peking
  University, Beijing 100871, China \\  
2. Collaborative Innovation Center of Quantum Matter, Beijing 100871, China}

\begin{abstract}
A self-consistent field theory is introduced and used to investigate the
thermodynamics and spin dynamics of an $S = 1$ quantum spin system 
with a magnetic Skyrmion. The temperature dependence of the Skyrmion
profile as well as the phase diagram are calculated. It is shown that
the Skyrmion carries a phase transition to the ferromagnetic phase of
first order with increasing temperature, while the  magnetization of
the surrounding ferromagnet undergoes a phase transition of second
order when changing to the paramagnetic phase. Furthermore, the 
electric field driven annihilation process of the Skyrmion is described
quantum mechanical by solving the time dependent Schr\"odinger
equation. The results are compared with the trajectories of the semi-classical
description of the spin expectation values using a differential equation
similar to the classical Landau-Lifshitz-Gilbert equation.   
\end{abstract}

\maketitle

\section{Introduction}
Magnetic Skyrmions have undergone an intensive attention during the
last years caused by the ideas to use them for data storage and as
part of logic devices
\cite{zhangSR15III,zhouNatComm14,fertNatNano13,zhangSR15,zhangSR15II}. The 
advantage of magnetic Skyrmions is the stability due to their
topology. Secondly, magnetic Skyrmions are smaller than other domain
wall formations and therefore a higher storage density and a faster
information flow are possible \cite{fertNatNano13}. Furthermore,
magnetic Skyrmions can be found in magnetic film systems at the 
atomic length \cite{rommingScience13,barkerPRL16} scale as well as in
magnetic films with  micrometer dimensions
\cite{jiangSCIENCE15,wooNatMat16etal,muhlbauerSCIENCE09}. This makes it 
possible to design devices on several length scales or to combine
these scales.  

So far all computer simulations describing the dynamics of magnetic
Skyrmions using a classical description. Furthermore there are only a
hand full theoretical studies investigating the thermodynamics of
Skyrmions. However, Usov et al. \cite{usovJMMM04} and Nieves et
al. \cite{nievesPRB14} have pointed out that a classical description
is not always convenient. Within this publication a quantum mechanical
self-consistent field (SCF) theory using a Hartree-Fock approximation
is used to describe the thermodynamics and after further modification
to describe the spin dynamics of a single Skyrmion.

SCF theories in general offer the chance to investigate complex
many-body systems by reducing them to local one-body problems. This
fact together with the dispelling of the fluctuations make these
theories successful. So far, several theories like the Ginzburg-Landau
theory \cite{stanleyBOOK71,coeyBOOK09}, the Stoner model of itinerant
magnets \cite{noltingBOOKMagE}, the random phase approximation (RPA)
in the description of many-body Green's functions
\cite{noltingBOOKMagE}, the Hartree-Fock theory
\cite{LevineBOOK91,pielaBOOK07}, and some other theories have been
proposed using the same methodology. 

The within this study used quantum mechanical SCF theory helps to
increase the system size which can be addressed. Actually it is
possible via exact diagonalization to describe the dynamics and
thermodynamics of 60 quantum spins with spin quantum number $S = 1/2$
on a square lattice. With the SCF theory the description becomes local
and entanglement among the spins plays no role. Therefore,
the Hilbert space and the corresponding matrices of the system are drastically  
reduced and larger system sizes can be addressed: within
this publication e.g. $200 \times 200 = 40000$ quantum spins $S = 1$
on a quadratic lattice. However, the used theory makes sense only if
entanglement is negligible. In systems with strong entanglement the
SCF theory will lead to false results.  

The publication is organized as follow: In Sec.~\ref{s:model} the quantum
mechanical SCF Hamilton operator $\hat{\mathrm{H}}$ is
introduced. Sec.~\ref{s:thermo} describes the temperature dependence   
of the local magnetization inside and outside the Skyrmion. The eigenvalue
and eigenfunction of the ground state of the Hamiltonian
$\hat{\mathrm{H}}$ are discussed in Sec.~\ref{s:EF} and
Sec.~\ref{s:dynamics} describes the switching dynamics of the Skyrmion
driven by an electric field oriented perpendicular to the film
plane. The publication ends with a summary (Sec.~\ref{s:summary}). 

\section{Model} \label{s:model}
The quantum mechanical SCF theory using a Hartree-Fock
approximation is used to describe a single magnetic Skyrmion. 
So far, nearly all studies investigating magnetic Skyrmions using the
classical Heisenberg model or the micromagnetic approximation of the
Heisenberg model. However, the classical approach is not always
flexible enough to study the magnetic properties
\cite{usovJMMM04,nievesPRB14}. The description within this 
study is quantum mechanical and investigates a square lattice with
$200 \times 200$ quantum spins with spin quantum number $S = 1$. The
physical properties are described by the following Hamilton operator:  
\begin{eqnarray} \label{HamSCF}
\hat{\mathrm{H}} = &-&J \sum\limits_{n,m} \hat{\mathbf{S}}_n
\cdot \langle \hat{\mathbf{S}}_m \rangle  
-  \sum\limits_{n,m} {\boldsymbol{\cal D}}_{nm} \cdot (\hat{\mathbf{S}}_n
\times \langle \hat{\mathbf{S}}_m \rangle) \nonumber \\
&-& g\mu_B \sum\limits_n \mathbf{B} \cdot \hat{\mathbf{S}}_n \;.
\end{eqnarray}
$\hat{\mathrm{H}}$ describes the interaction of the spin operator
$\hat{\mathbf{S}}_n$ with the surrounding spins by their expectation
values $\langle \hat{\mathbf{S}}_m \rangle$. Therefore, the
description is reduced to a description of local spins. 

\begin{figure*}
\vspace{1mm} % ?????
\includegraphics*[width=5.75cm,bb = 210 385 585 680]{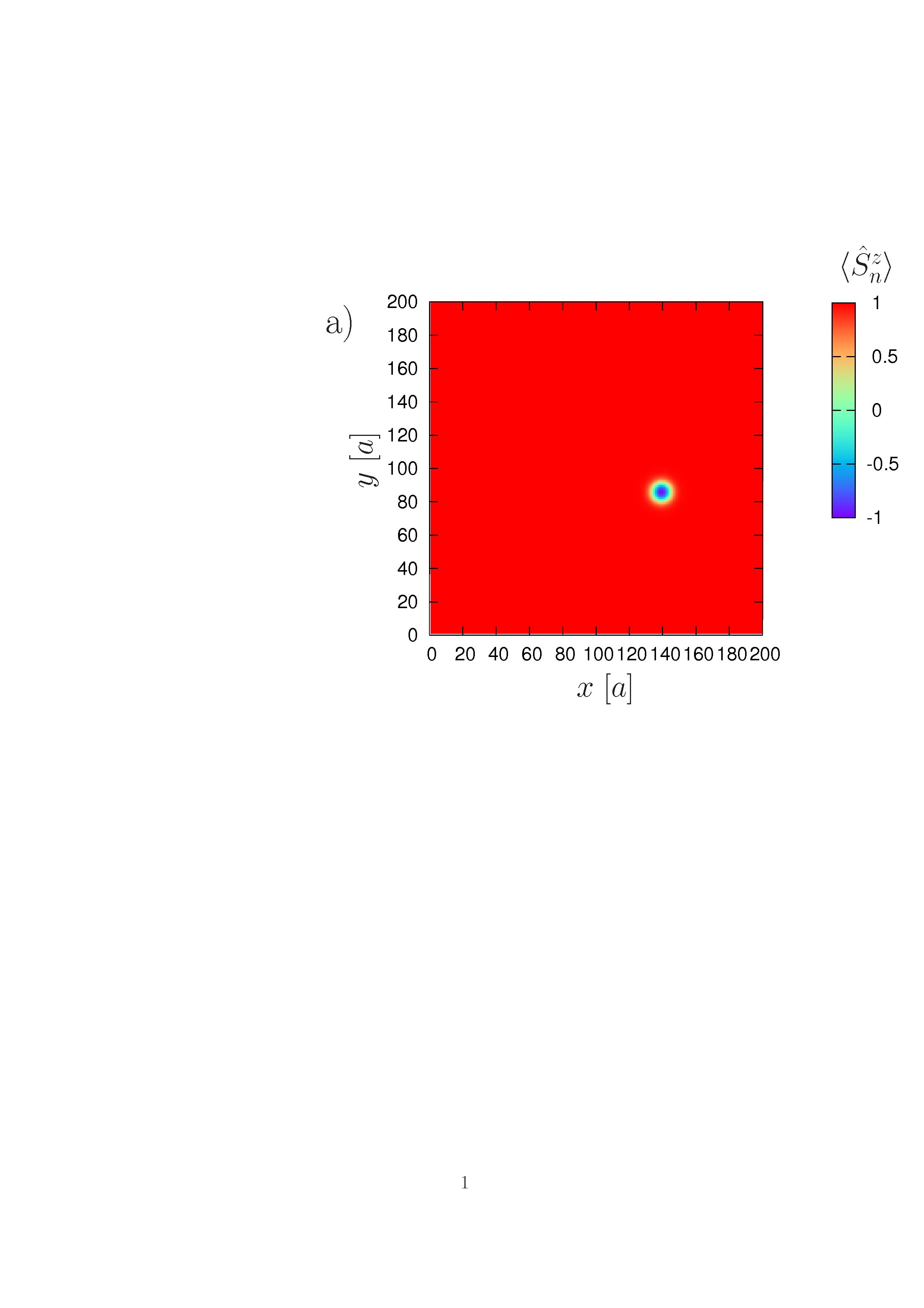}
\includegraphics*[width=4.5cm,bb = 65 255 540 710]{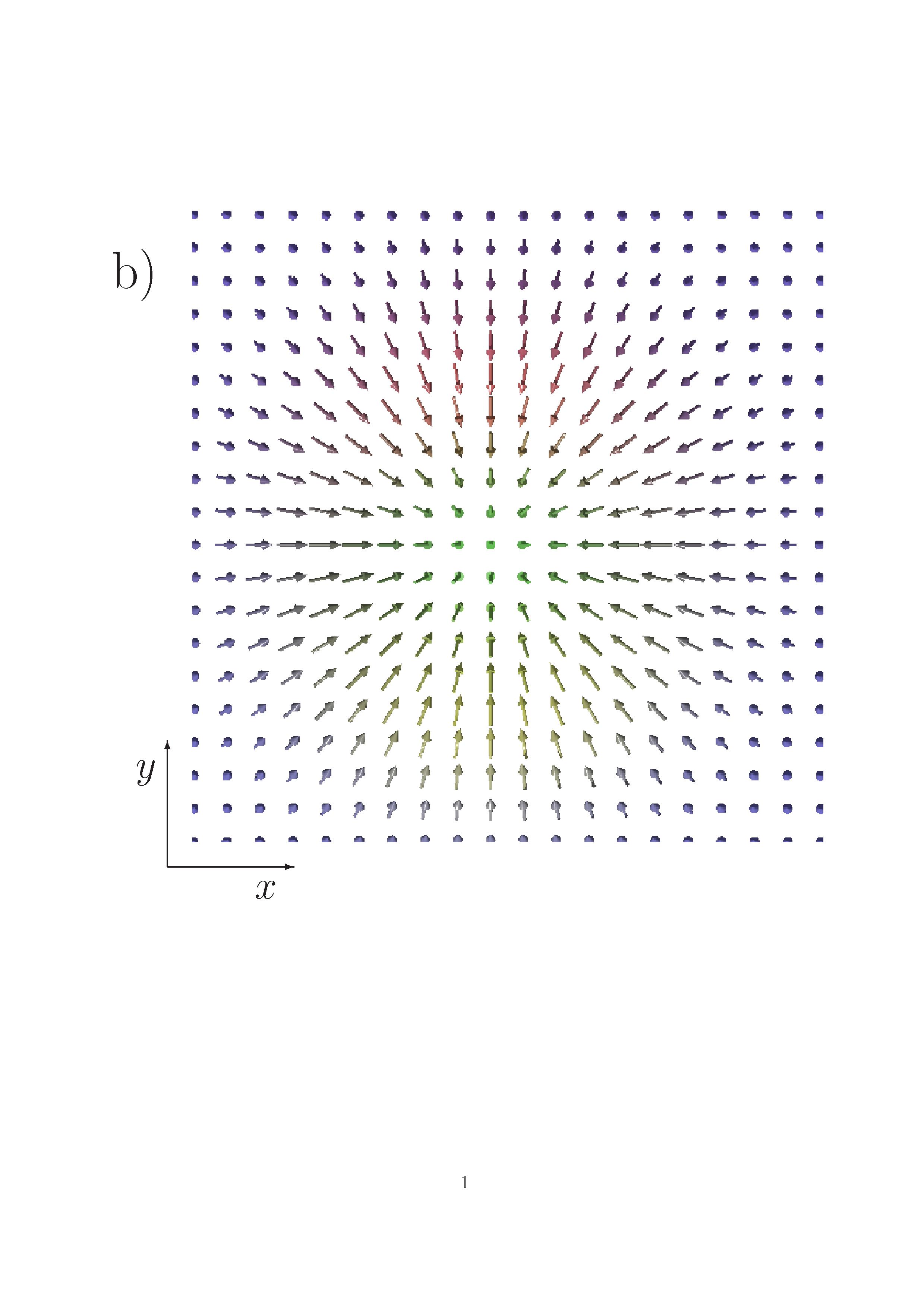}
\\
\includegraphics*[width=5.75cm,bb = 75 460 505 770]{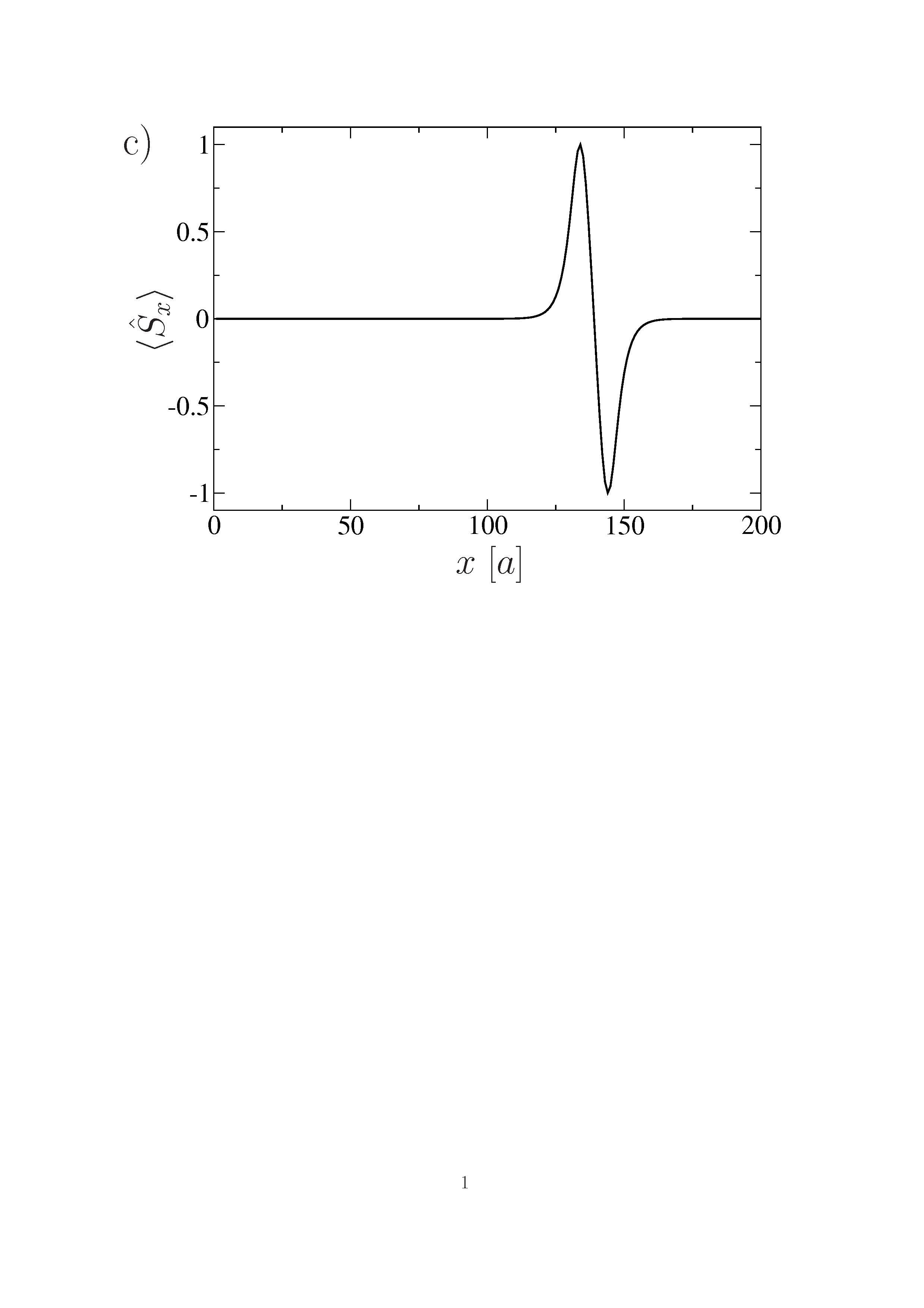}
\includegraphics*[width=5.75cm,bb = 75 460 505 770]{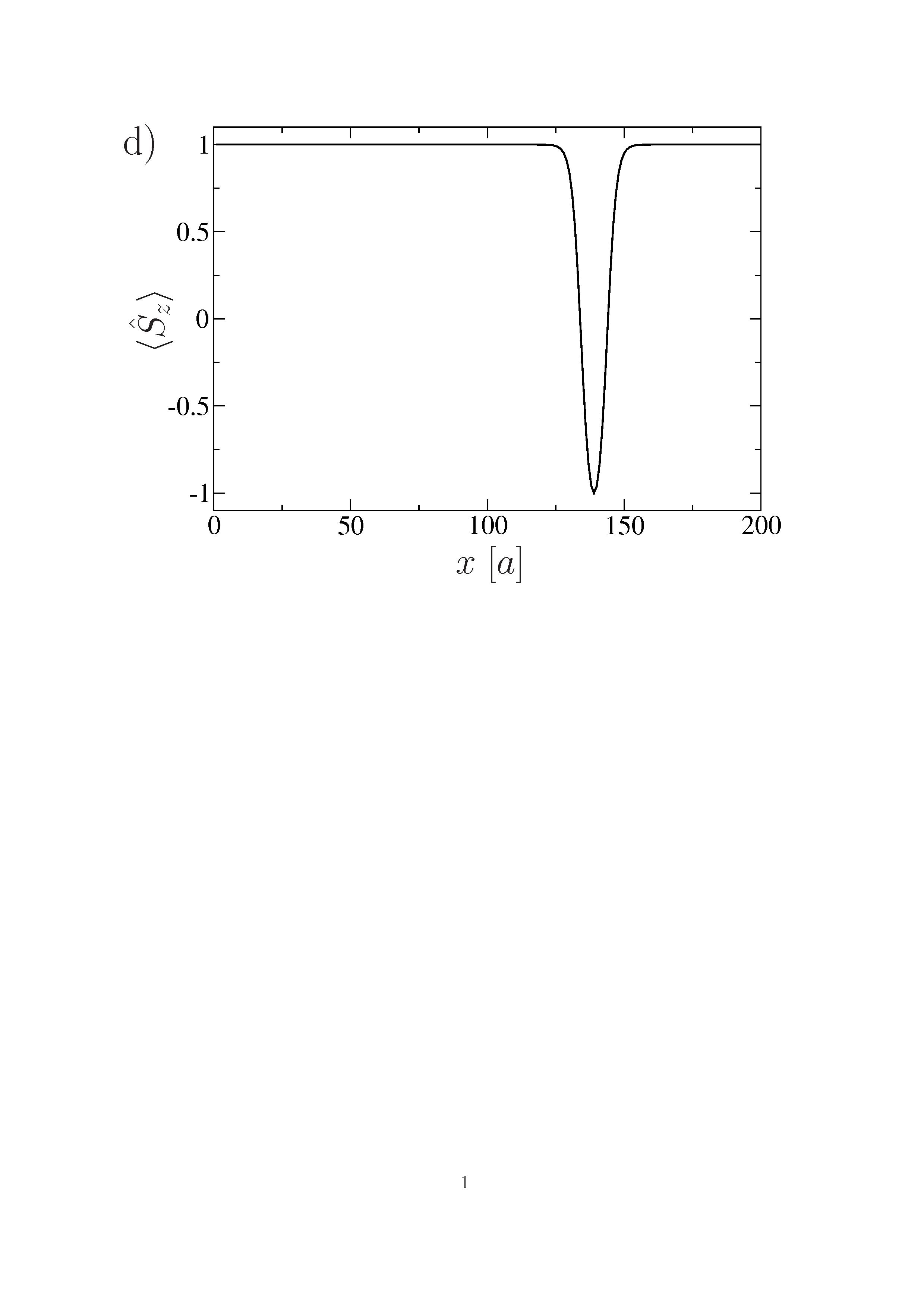}
  \caption{(color online) Single magnetic Skyrmion: a) $z$-component,
    b) in plane Hedgehog structure, c) and d) corresponding 
    profiles provided by a cut through the center of the Skyrmion. The
    Skyrmion structure and profiles are calculated with Eq.~(\ref{SCF})
    at $B = 4.5$ T.  
}       
  \label{f:pic1}
\end{figure*}

The first and second terms of the Hamilton operator $\hat{\mathrm{H}}$
describe the nearest neighbor exchange and Dzyaloshinsky-Moriya
interaction. Depending on the sign the exchange interaction is either
ferromagnetic $J > 0$ or antiferromagnetic $J < 0$. In the following a
ferromagnetic interaction is assumed. Together with the
exchange interaction the Dzyaloshinsky-Moriya interaction (DMI) lead
to a spin-spiral. The ${\boldsymbol{\cal D}}_{nm}$ are the DMI vectors
responsible for the direction and sense of rotation of the spiral 
\cite{dzyaloshinskyJPCS58,crepieuxJMMM98}. Within this publication the
${\boldsymbol{\cal D}}_{nm}$ are assumed to be in-plane (magnetic
film) vectors oriented perpendicular to the connection between two
nearest neighbor lattice sites: ${\boldsymbol{\cal D}}_{nm} \perp
\mathbf{r}_{nm}$. The third term in $\hat{\mathrm{H}}$
describes the coupling of the spins to an external magnetic field
(Zeeman term). Within this term $g$ is the Land\'e factor, $\mu_B$ the
Bohr magneton, and $\mathbf{B}$ the magnetic field. In the
following an external magnetic field perpendicular to the film plane
($z$-direction) is assumed to stabilize the Skyrmion. The resulting
Skyrmion can be seen in Fig.~\ref{f:pic1}. The values 
for the the ferromagnetic exchange interaction, the DMI and 
external field used in this study are $J = 7$ meV, ${\cal D}_{nm} =
2.2$ meV, and $B = 4.5$ T.    
\begin{figure*}
\vspace{1mm} % ?????  
\includegraphics*[width=5.cm,bb = 75 460 505 770]{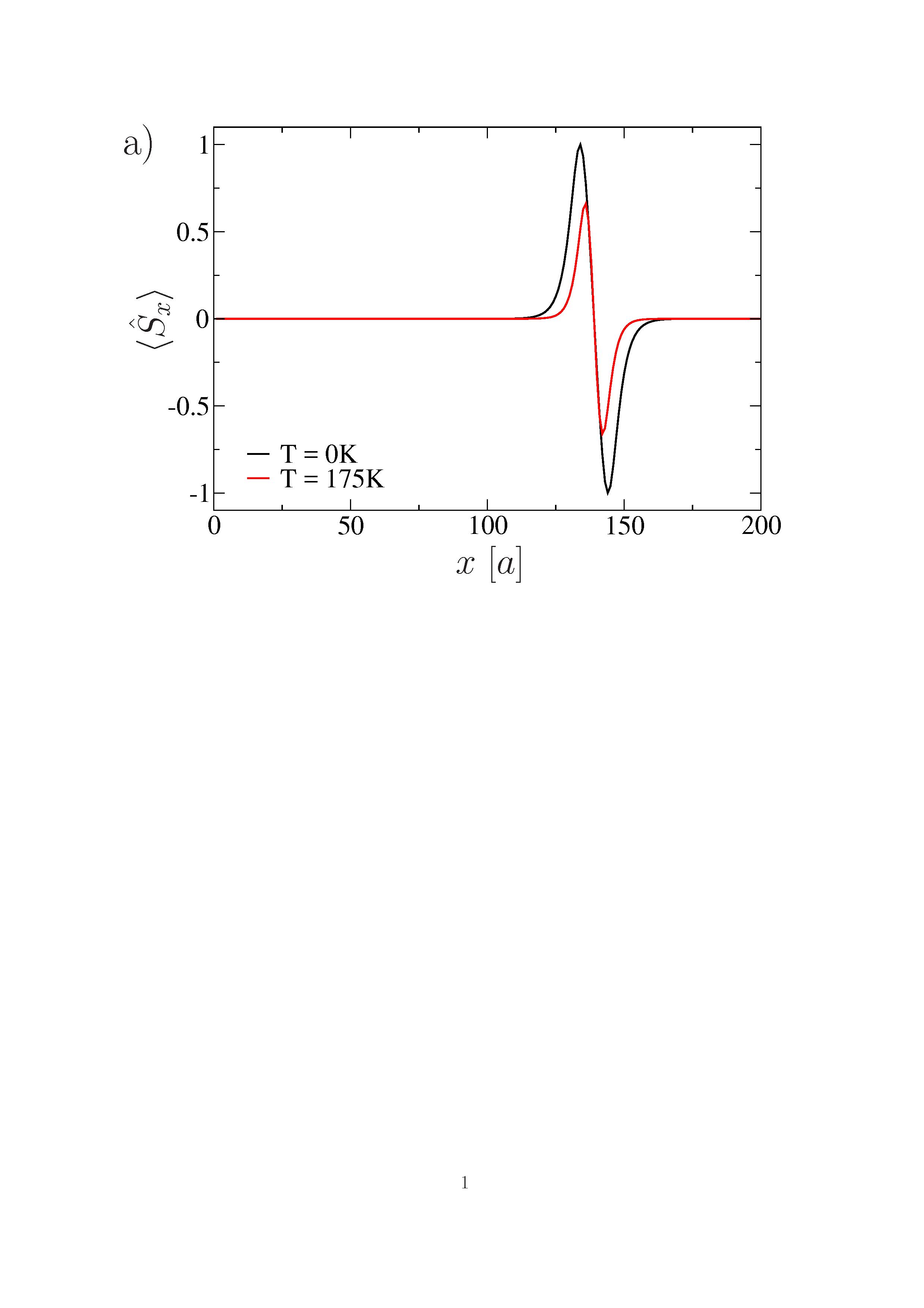} 
\includegraphics*[width=5.cm,bb = 75 460 505 770]{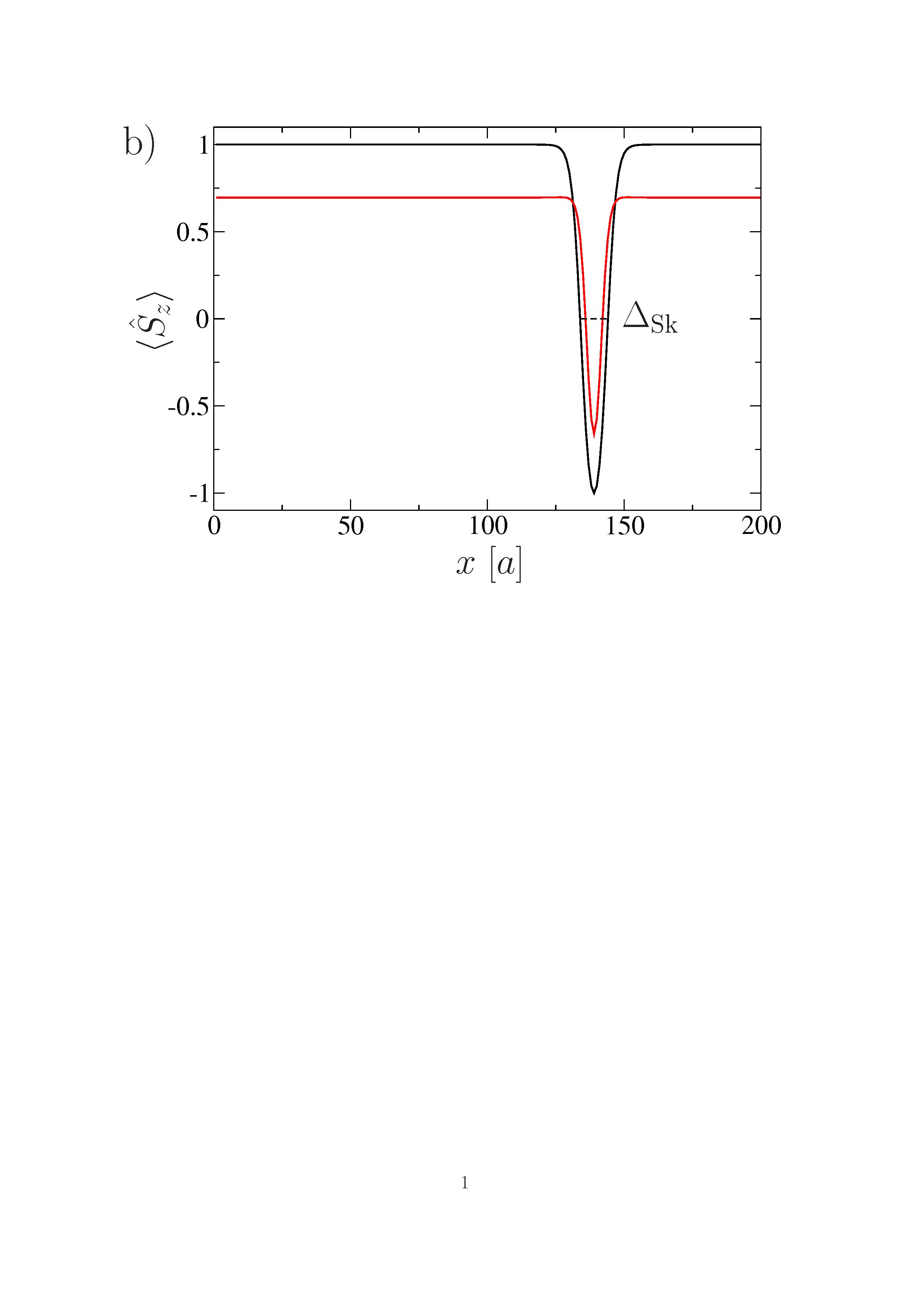} 
\includegraphics*[width=5.cm,bb = 75 455 505 770]{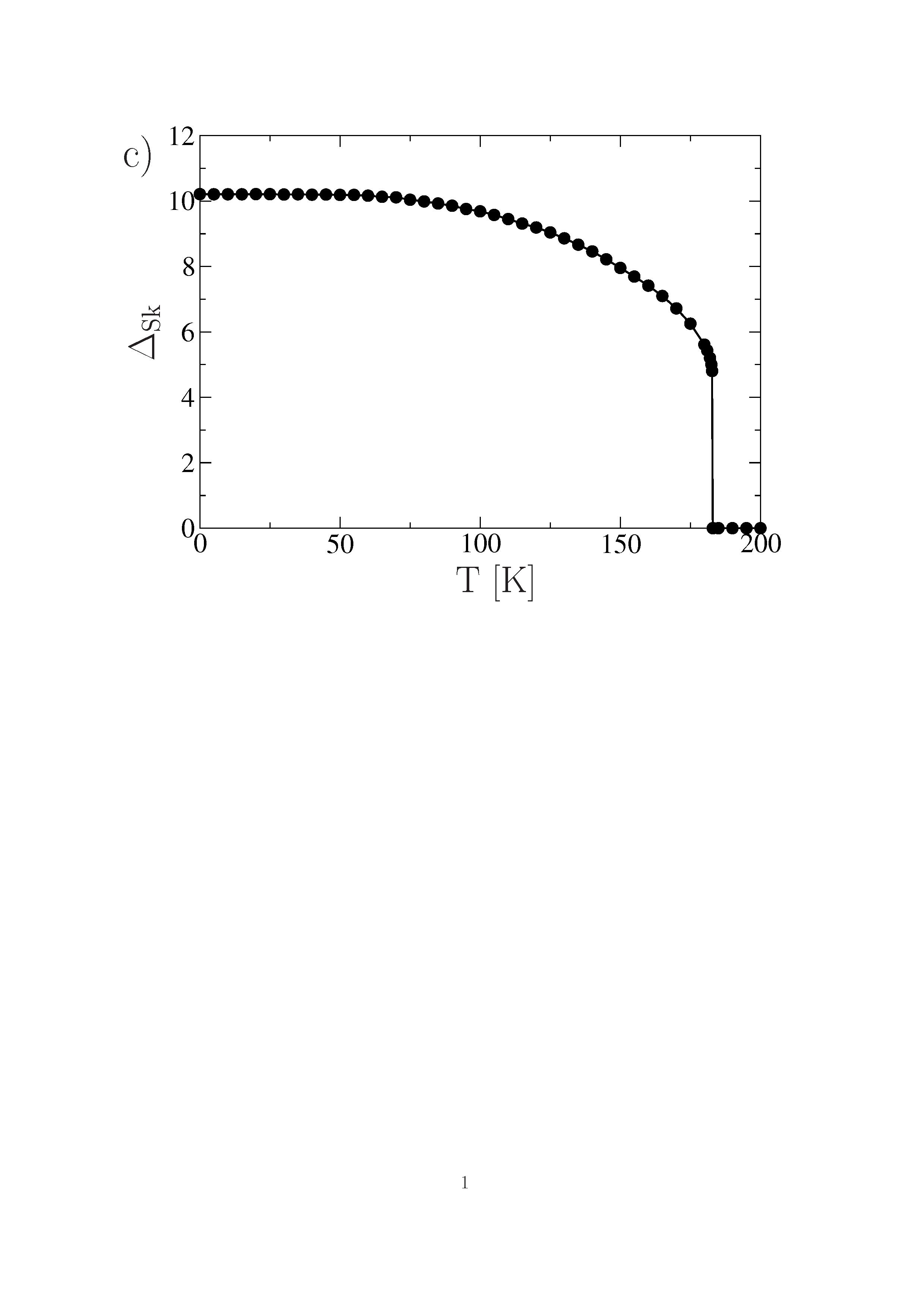} 
  \caption{(color online) Magnetic Skyrmion temperature dependence: a)
    and b) Skyrmion profiles at $T = 0$ K and $T = 175$ K, c)
    Skyrmion width as function of temperature.  
}       
  \label{f:pic2}
\end{figure*}
\begin{figure*}
\vspace{1mm} % ?????
\includegraphics*[width=5.5cm,bb = 75 455 505 770]{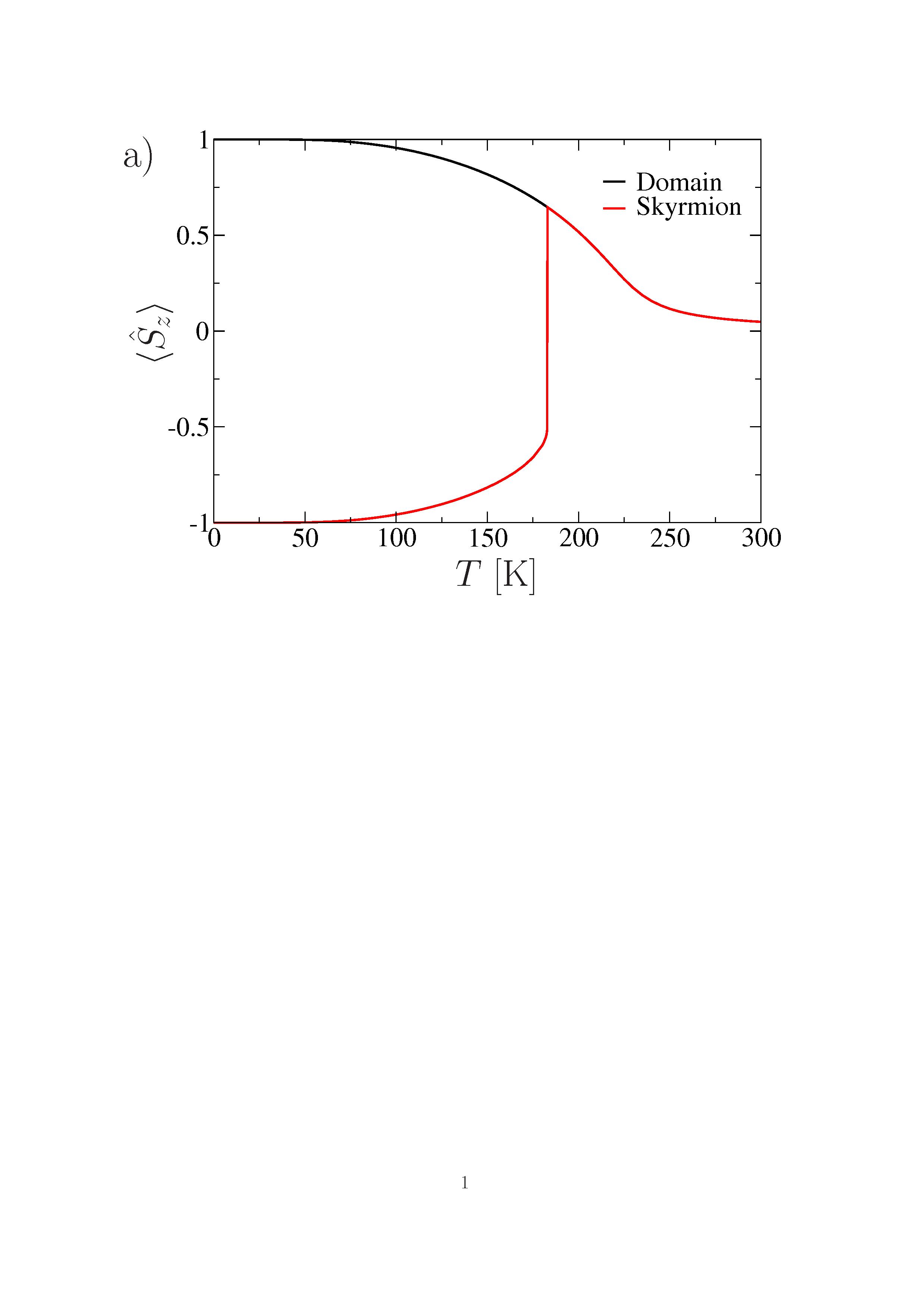}
\includegraphics*[width=5.5cm,bb = 75 455 505 770]{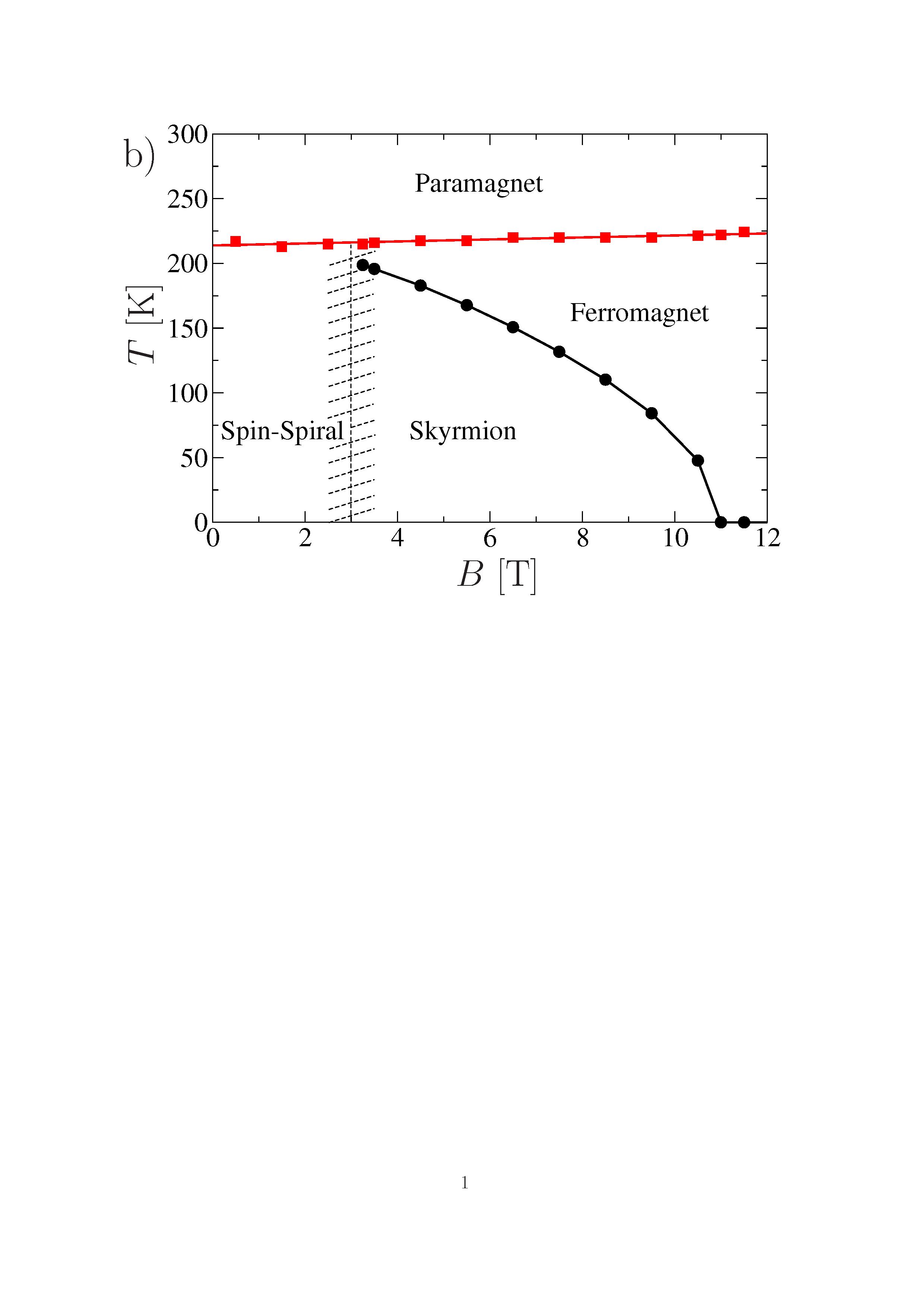}
  \caption{(color online) Temperature dependence: a) spin expectation
    values $\langle \hat{S}_z \rangle$ as function of temperature. The
    spin expectation values are calculated for the center of the Skyrmion
    and the surrounding domain. b) corresponding phase diagram 
    with first (Skyrmion $\leftrightarrow$ Ferromagnet) and second
    order (Skyrmion / Ferromagnet $\leftrightarrow$ Paramagnet) phase
    transitions. The dashed area is the transition area between the
    Skyrmionic and the ferromagnetic phase. The pictures correspond to
    $B = 4.5$ T.   
}       
  \label{f:pic3}
\end{figure*}

With the definition of the effective field
$\mathbf{B}_n^{\mathrm{eff}} =  -\frac{\partial
\hat{\mathrm{H}}}{\partial \hat{\mathbf{S}}_n}$, where
$\mathbf{B}_n^{\mathrm{eff}} = (B_n^x,B_n^y,B_n^z)^T$ is a real
number, the Hamilton operator
$\hat{\mathrm{H}}$ can be written as:     
\begin{eqnarray} \label{HamSCF2}
\hat{\mathrm{H}} = \sum\limits_n
\mathbf{B}_n^{\mathrm{eff}} \cdot \hat{\mathbf{S}}_n \;. 
\end{eqnarray}
Eq.~(\ref{HamSCF2}) already shows the local character of the Hamiltonian. 

Then, with the definition: 
\begin{equation}
\hat{\mathrm{h}}_n = \mathbf{B}_n^{\mathrm{eff}} \cdot
\hat{\mathbf{S}}_n \;,
\end{equation}
$\hat{\mathrm{H}}$ can be finally written as sum of the local
Hamilton operators $\hat{\mathrm{h}}_n$:
\begin{eqnarray} \label{HamSCF3}
\hat{\mathrm{H}} = \sum\limits_n \hat{\mathrm{h}}_n 
\;.
\end{eqnarray}

\section{Thermodynamical studies} \label{s:thermo}

The temperature dependence of the ground state configuration of a spin
system, like a magnetic Skyrmion, can be easily investigated. From
statistical physics it is known that the spin expectation values are
defined by:   
\begin{eqnarray}
\langle \hat{\mathbf{S}}_n \rangle =
\frac{\mathrm{Tr}\left(\hat{\mathbf{S}}_n \exp(\beta 
  \hat{\mathrm{H}})\right)}{\mathrm{Tr}\left(
  \exp(\beta \hat{\mathrm{H}})\right)} \;.
\end{eqnarray}
As usual $\beta$ is the inverse temperature: $\beta = \frac{1}{k_BT}$. 
In case of a quantum spin system the trace is the sum over all quantum
numbers and therefore we find after some algebra an equation which can
be easily solved numerical:
\begin{eqnarray} \label{SCF}
\langle \hat{\mathbf{S}}_n \rangle = g \mu_B {\cal B}_S(\beta
B_n^{\mathrm{eff}}) \;.
\end{eqnarray}
Within this equation ${\cal B}_S(\beta
B_n^{\mathrm{eff}})$ is the Brillouin function:
\begin{eqnarray}
{\cal B}_S(\beta B_n^{\mathrm{eff}}) =
\frac{2S+1}{2S}\mathrm{coth}\left(\frac{2S+1}{2S}\beta B_n^{\mathrm{eff}}\right)
-\frac{1}{2S}\mathrm{coth}\left(\frac{1}{2S}\beta B_n^{\mathrm{eff}} \right) \;.
\end{eqnarray}

Fig.~\ref{f:pic1} shows the Skyrmionic configuration at zero
temperature ($T = 0$ K) and $B = 4.5$ T which can be derived
by solving Eq.~(\ref{SCF}) self-consistent for a random initial
configuration or a given Skyrmionic structure. Such a Skyrmion
configuration can be found for $S = 1$, but also for $S = 1/2$ or in
the case of a classical spin system ($S = \infty$). In
Fig.~\ref{f:pic1}a) the whole system ($z$-component of the spin
expectation value $\langle \hat{\mathbf{S}} \rangle$) can be seen with 
a single Skyrmion located at $x = 139$ $a$, $y = 86$ $a$, where $a$ is the
lattice constant of the system. Fig.~\ref{f:pic1}b) provides the
microscopic structure of the Skyrmion which due to the in plane DMI
vectors is a Hedgehog structure, meaning all magnetic moments 
pointing to the center of the Skyrmion. Fig.~\ref{f:pic1}c) and
Fig.~\ref{f:pic1}d) give the Skyrmion profiles which appear for a
cut through the middle of the Skyrmion ($y = 86$ $a =$
const.). Similar profiles can be found also using other simulation
methods like Langevin spin dynamics \cite{garciaPRB98} or Monte Carlo
simulations \cite{hagemeisterNatComm15}, but as already said the
advantage of the SCF theory lays in the fact that temperature
effects can be investigated without interfering
noise. Fig.~\ref{f:pic2} shows the Skyrmion profiles for $T = 0$ K and 
$T = 175$ K. As expected the magnetization decreases with increasing
temperature, but  the profile of the Skyrmion does not changes even if
the Skyrmion shrinks. The reduction of Skyrmion size with increasing
temperature can be manifested with aid of the Skyrmion radius
$\Delta_{\mathrm{Sk}}$. In the following, $\Delta_{\mathrm{Sk}}$ shall be 
defined as the distance between the two zero crossings of $\langle
\hat{S}_z \rangle$ [see Fig.~\ref{f:pic2}b)]. $\Delta_{\mathrm{Sk}}$ as
function of temperature $T$ is given in Fig.~\ref{f:pic2}c),
$\Delta_{\mathrm{Sk}}$ decreases with increasing
temperature and becomes zero at the critical temperature
$T_C^{\mathrm{Sk}} = 183$ K ($B_z = 4.5$ T). Above this 
temperature the ferromagnetic state is the ground state, meaning
$\Delta_{\mathrm{Sk}} = 0$. The abrupt change of
$\Delta_{\mathrm{Sk}}$ indicates that the transition is of 
first order which means that during the annihilation of the Skyrmion
energy is released.   

The first order phase transition can be observed also by investigating the local
magnetization (spin expectation value $\langle \hat{S}_n^z \rangle$) as
function of temperature $T$. More precisely the first order
transition can be seen if the magnetization inside the Skyrmion is
investigated. In the case of the surrounding ferromagnetic environment the
phase transition is of second order. In Fig.~\ref{f:pic3}a) the spin
expectation values $\langle \hat{S}_n^z \rangle$ correspond to the
center of the Skyrmion at $x = 139$ $a$, $y = 86$ $a$ and the
surrounding ferromagnetic domain: $x = 80$ $a$, $y = 80$ $a$ are
plotted as function of temperature. In both
cases the absolute value $|\langle \hat{S}_n^z \rangle|$ are equal to
one at zero temperature and decrease with increasing temperature. A careful
analysis of the magnetization shows that inside the Skyrmion $|\langle
\hat{S}_n^z \rangle|$ decreases faster than in the domain. In
Fig.~\ref{f:pic3}a) $\langle \hat{S}_n^z \rangle$ is plotted instead
of $|\langle \hat{S}_n^z \rangle|$ to show the opposite orientation of
$\langle \hat{S}_n^z \rangle$ inside the Skyrmion and the surrounding
domain. The different temperature dependences of the magnetization
inside and outside the Skyrmion are similar to the behavior of the
magnetization of a magnetic Vortex \cite{wieserPRB06} however with
one significant difference. In the case of the Vortex the phase
transitions of the magnetization of the Vortex core and the
surrounding domain are both of second order and not of first and
second order. Nevertheless, in both cases (Skyrmion and magnetic
Vortex) we find two different critical temperatures
$T_C^{\mathrm{Sk}}$ (Skyrmion) and $T_C^{\mathrm{FM}}$ 
(surrounding Ferromagnet): $T_C^{\mathrm{Sk}} = 183$ K and
$T_C^{\mathrm{FM}} = 217.5$ K. Remark: These values correspond to an external
magnetic field with $B_z = 4.5$ T. In general $T_C^{\mathrm{Sk}}$ and
$T_C^{\mathrm{FM}}$ depend on the external magnetic field. The
dependence and therefore the phase diagram is given in
Fig.~\ref{f:pic3}b). As said before, the transition from the 
Skyrmion phase to the ferromagnetic phase is a first order phase
transition and the phase transition from the ferromagnetic phase to
the paramagnetic phase is of second order. Furthermore, there is
another transition from the spin-spiral phase to the Skyrmion phase
with increasing external field. The striped area marks the transition
area where both phases coexist.

\section{Eigenfunctions of $\hat{\mathrm{H}}$} \label{s:EF}
Due to the fact that $\hat{\mathrm{H}}$
[Eq.~(\ref{HamSCF})-(\ref{HamSCF3})] is a quantum mechanical operator
respectively mathematically a matrix it is possible to calculate the
corresponding eigenvalues and eigenfunctions. In Sec.~\ref{s:model} it
has been shown that the Hamiltonian $\hat{\mathrm{H}}$
can be written as sum of the local Hamilton operators
$\hat{\mathrm{h}}_n$. Therefore, the problem reduces to calculate the
eigenvalues $E_n$ and eigenvectors $\phi_n$ of $\hat{\mathrm{h}}_n$:
\begin{equation}  \label{eigen}
\hat{\mathrm{h}}_n {\boldsymbol \phi}_n = E_n {\boldsymbol \phi}_n \;.
\end{equation}
Finally, the eigenvalues and eigenvectors or eigenfunctions of
$\hat{\mathrm{H}}$ are the product states of $\phi_n$
respectively the sums over $E_n$.  
\begin{figure*}
\vspace{1mm} % ?????
\includegraphics*[width=6.25cm,bb = 225 395 585 680]{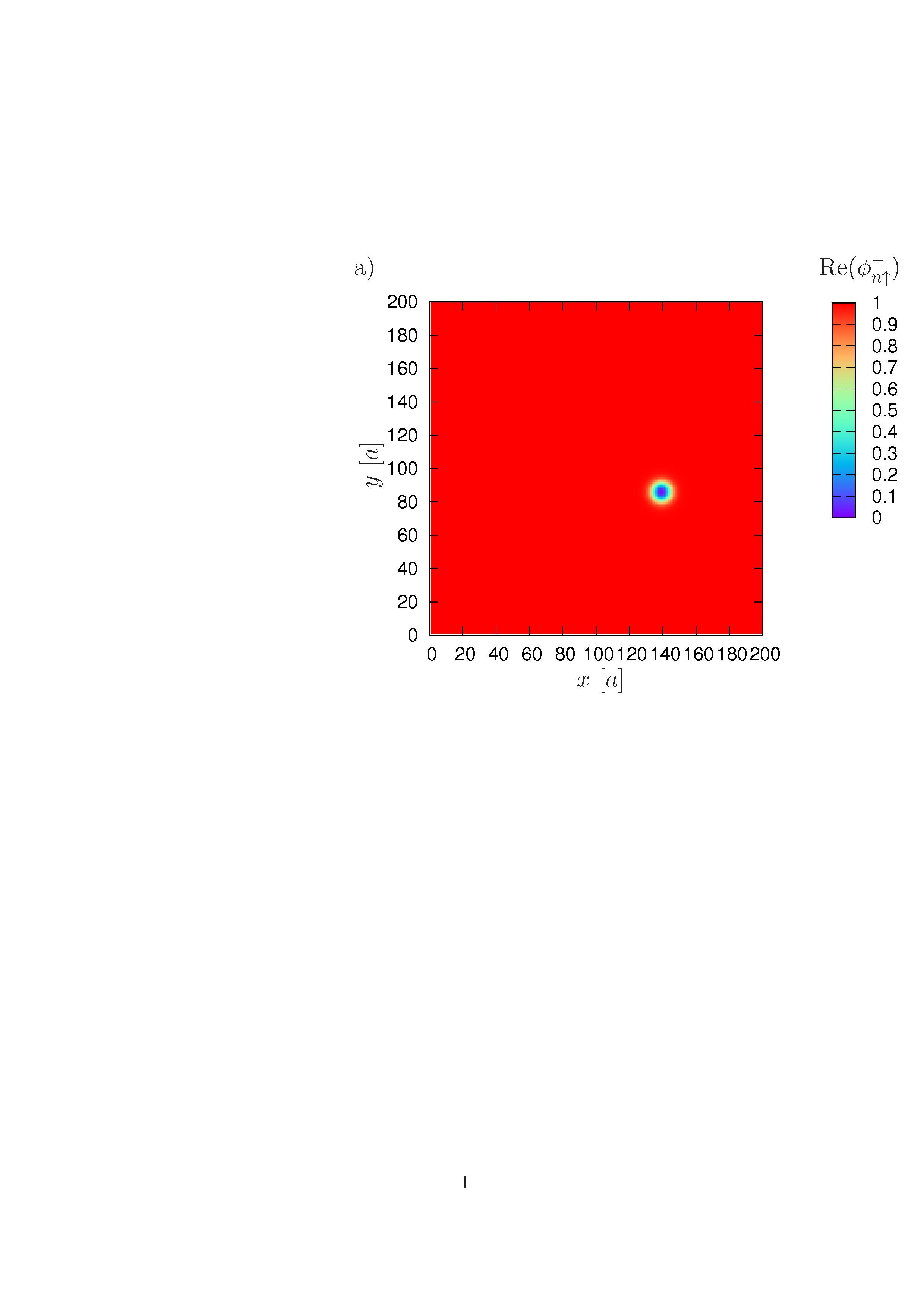}
\hspace{5mm}
\includegraphics*[width=6.25cm,bb = 225 395 585 680]{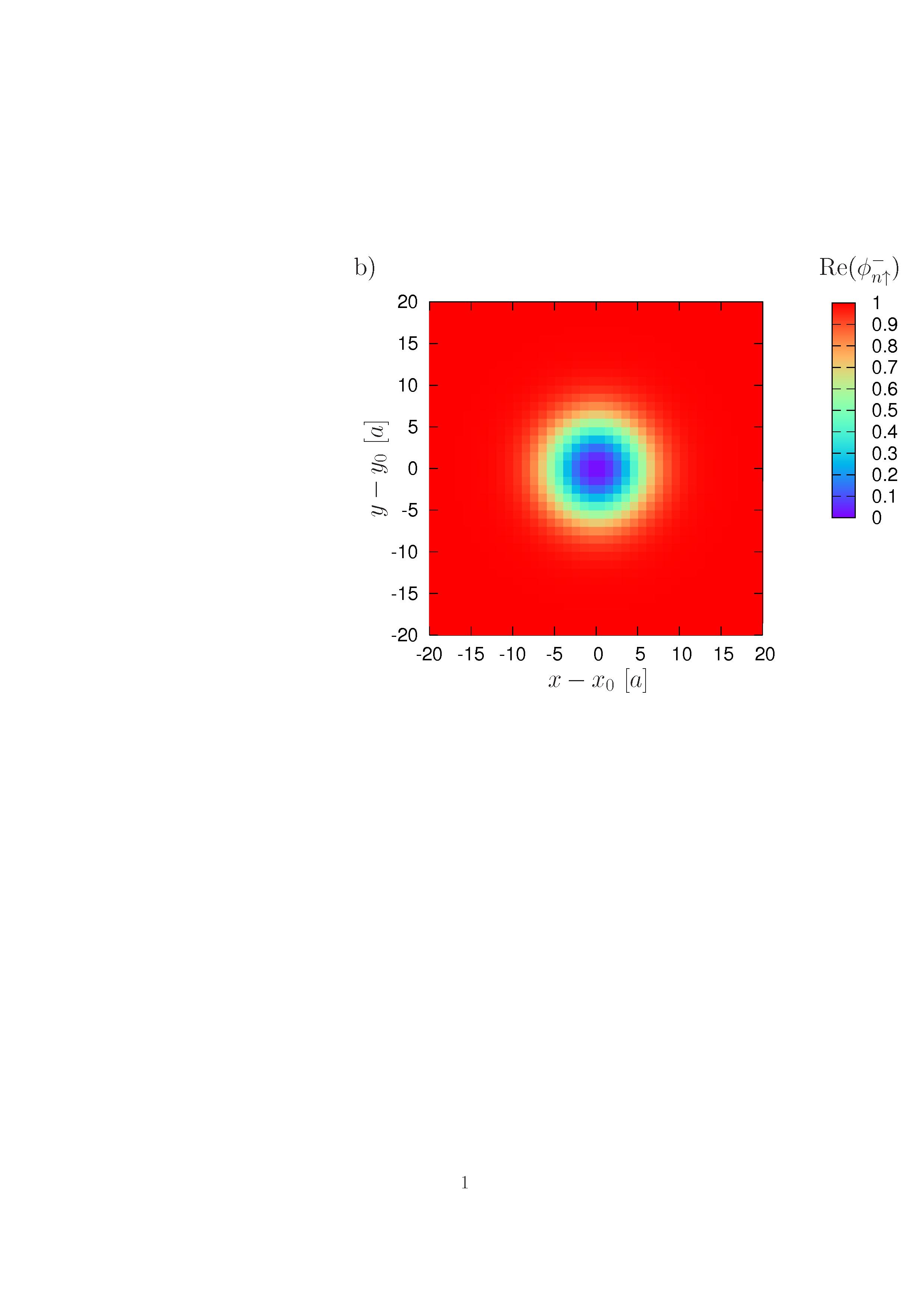}
\includegraphics*[width=6.25cm,bb = 225 395 585 680]{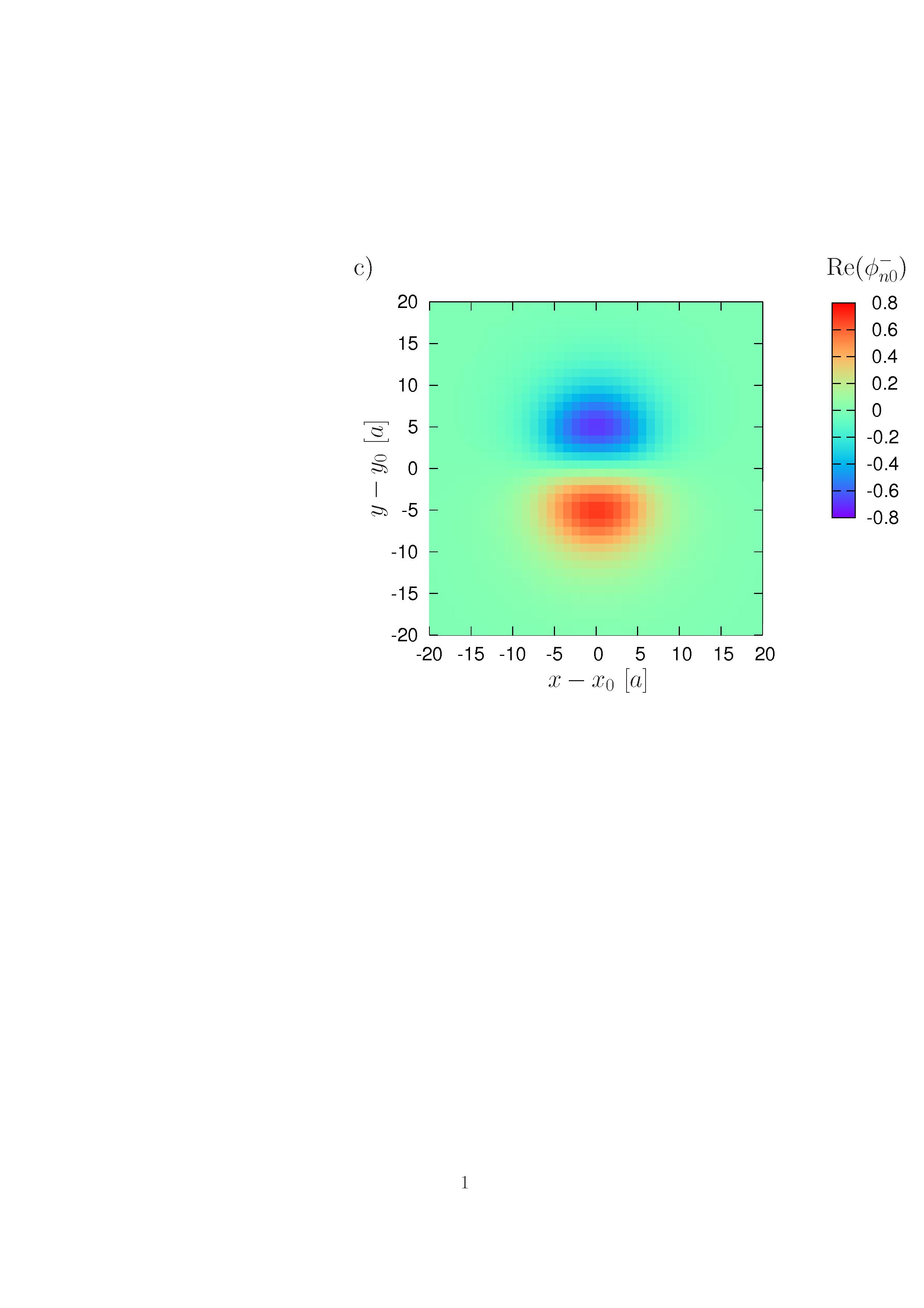}
\hspace{5mm}
\includegraphics*[width=6.25cm,bb = 225 395 585 680]{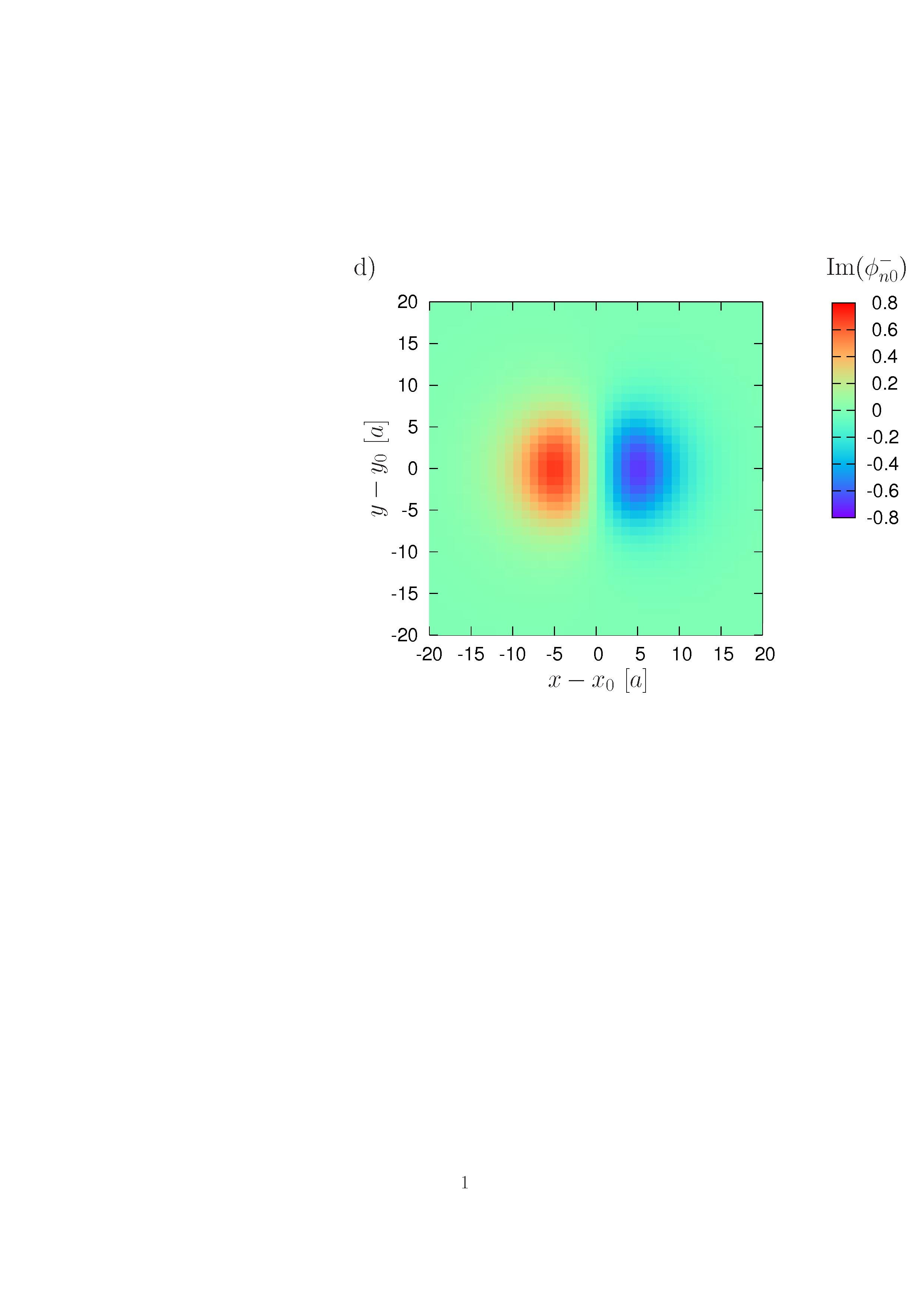}
\includegraphics*[width=6.25cm,bb = 225 395 585 680]{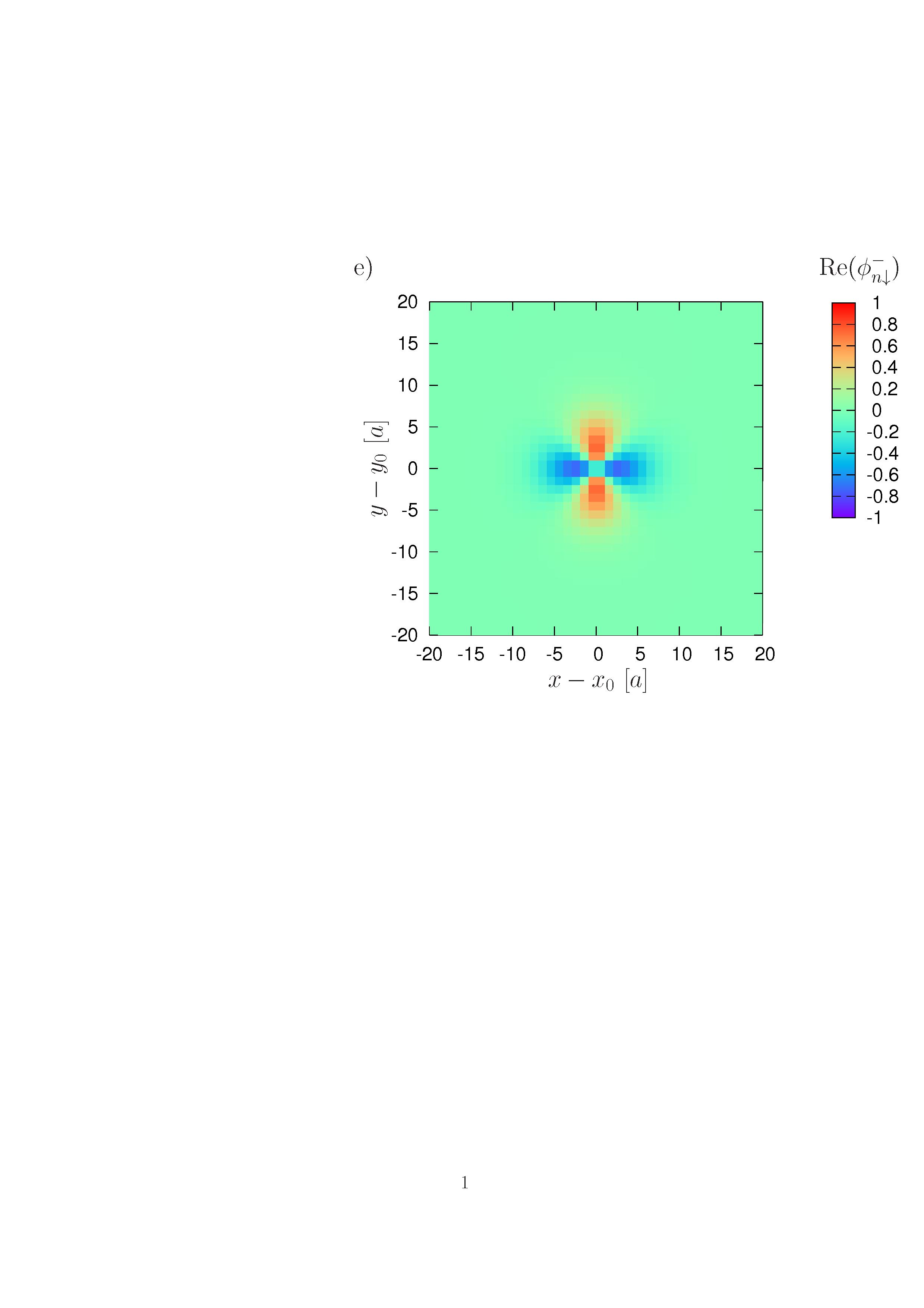}
\hspace{5mm}
\includegraphics*[width=6.25cm,bb = 225 395 585 680]{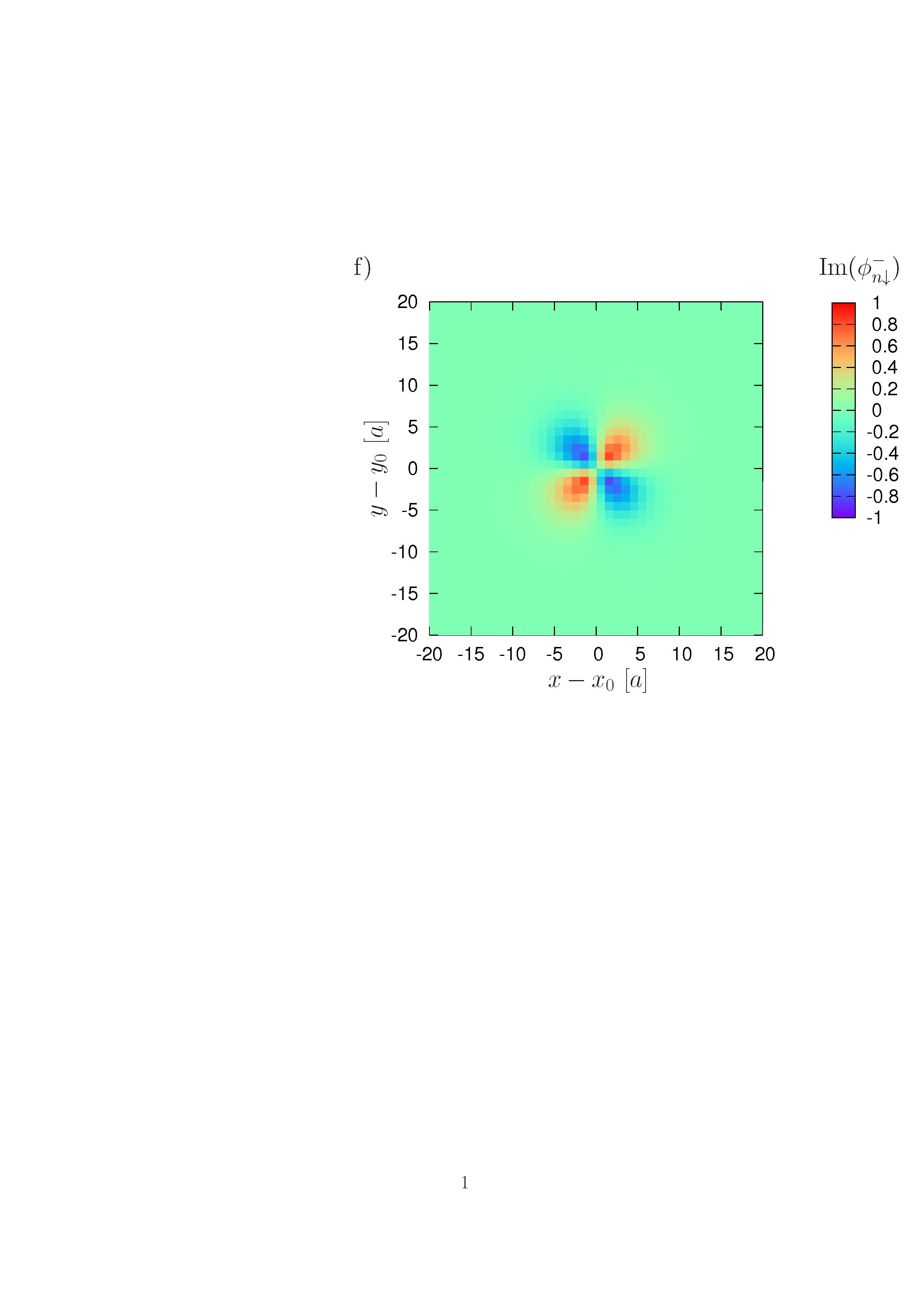}
  \caption{(color online) Eigenfunction of the ground state of a
    magnetic Skyrmion with $S = 1$. 
}       
  \label{f:pic4}
\end{figure*}

Mathematically, Eq.~(\ref{eigen}) is a matrix equation with the
Hamilton operator $\hat{\mathrm{h}}_n$:
%\begin{small}
\begin{eqnarray}
\hat{\mathrm{h}}_n = \left(\begin{array}{ccc} -B_n^z &
\frac{-B_n^x + i B_n^y}{\sqrt{2}} & 
0 \\ \frac{-B_n^x - i B_n^y}{\sqrt{2}} &
0 & \frac{-B_n^x + i B_n^y}{\sqrt{2}} \\
0 & \frac{-B_n^x - i B_n^y}{\sqrt{2}} & 
B_n^z \end{array} \right) 
\end{eqnarray}
and the eigenvectors:
\begin{eqnarray}
{\boldsymbol \phi}_n &=& \left(\begin{array}{c} \phi_{n\uparrow}
  \\ \phi_{n0} \\ \phi_{n\downarrow} \end{array}\right) \;.
\end{eqnarray}

%\end{small}
Using a standard diagonalization method the eigenenergies $E_n$ of
$\hat{\mathrm{h}}_n$ are easily calculated: 
\begin{eqnarray} \label{ResultEigenvalues}
E_n^{\pm} &=& \pm \mathrm{B}_n^{\mathrm{eff}} \\
E_n^0 &=& 0 \;.
\end{eqnarray}
Within Eq.~(\ref{ResultEigenvalues}) the effective fields
$\mathrm{B}_n^{\mathrm{eff}}$ are given by:
\begin{eqnarray}
\mathrm{B}_n^{\mathrm{eff}} = \sqrt{\left(B_n^x\right)^2 +
  \left(B_n^y\right)^2 + \left(B_n^z\right)^2} \;,
\end{eqnarray}
and the ground state is characterized by $E_n^- =
-\mathrm{B}_n^{\mathrm{eff}}$, while the corresponding eigenvector is:
\begin{eqnarray}
{\boldsymbol \phi}_n^- &=& \left(\begin{array}{c}
  \frac{\mathrm{B}_n^{\mathrm{eff}} 
    + B_n^z}{2\mathrm{B}_n^{\mathrm{eff}}}
  \\ \frac{\sqrt{2}}{2\mathrm{B}_n^{\mathrm{eff}}} \left(B_n^x + i
  B_n^y\right) \\ \left(\frac{\mathrm{B}_n^{\mathrm{eff}} -
  B_n^z}{2\mathrm{B}_n^{\mathrm{eff}}} \right)\!\!\!\left(\frac{B_n^x + i
    B_n^y}{B_n^x - i B_n^y}\right) \end{array}\right) \;.
\end{eqnarray}

Fig.~\ref{f:pic4} shows the real and imaginary parts of the vector components
$\{\phi^-_{n\uparrow}, \phi^-_{n0}, \phi^-_{n\downarrow}\} \in \mathbb{C}$ of the
eigenvector ${\boldsymbol \phi}^-_n$ corresponding to the Skyrmion shown in
Fig.~\ref{f:pic1}. Fig.~\ref{f:pic4}a) provides the total system while
Fig.~\ref{f:pic4}b)-f) are zoomed into the area of the Skyrmion. In
Fig.~\ref{f:pic4}a), the Skyrmion is clearly visible as spot in the
surrounding ferromagnetic environment which is characterized by
${\boldsymbol \phi}^-_n = (1,0,0)^T$, $n\not\in$. The Skyrmion itself
is described by all three components $\phi^-_{n\uparrow}$,
$\phi^-_{n0}$, and $\phi^-_{n\downarrow}$ of ${\boldsymbol \phi}^-_n$. 
Not shown in Fig.~\ref{f:pic4} is the imaginary part of
$\phi^-_{n\uparrow}$ which has been set to zero during the
calculation. 

\section{Quantum spin dynamics} \label{s:dynamics}

So far $\hat{\mathrm{H}}$ has been used to describe the
thermodynamics of the magnetic Skyrmion. But, the Hamilton operator
$\hat{\mathrm{H}}$ can also be used for the description of the spin
dynamics of the Skyrmion. The underlying equation of motion is the
time dependent Schr\"odinger equation with an additional damping term
\cite{wieserEPJB15}: 
\begin{eqnarray} \label{tdse1}
i\hbar (1-\lambda^2) \frac{\mathrm{d}}{\mathrm{d}t}|\Psi \rangle =
\hat{\mathrm{H}}|\Psi \rangle - i \lambda (
\hat{\mathrm{H}} - \langle
\hat{\mathrm{H}} \rangle )|\Psi \rangle \;.
\end{eqnarray} 
Within the Schr\"odinger equation $\lambda$ is a constant describing
the strength of the energy dissipation ($\lambda \geq
0$). Furthermore, $\langle \hat{\mathrm{H}} \rangle = \langle \Psi |   
\hat{\mathrm{H}} | \Psi \rangle$, and
$\hat{\mathrm{H}}$ given by Eq.~(\ref{HamSCF3}).   

\begin{figure*}
\vspace{1mm} % ?????  
\includegraphics*[width=5.75cm,bb = 60 450 520 770]{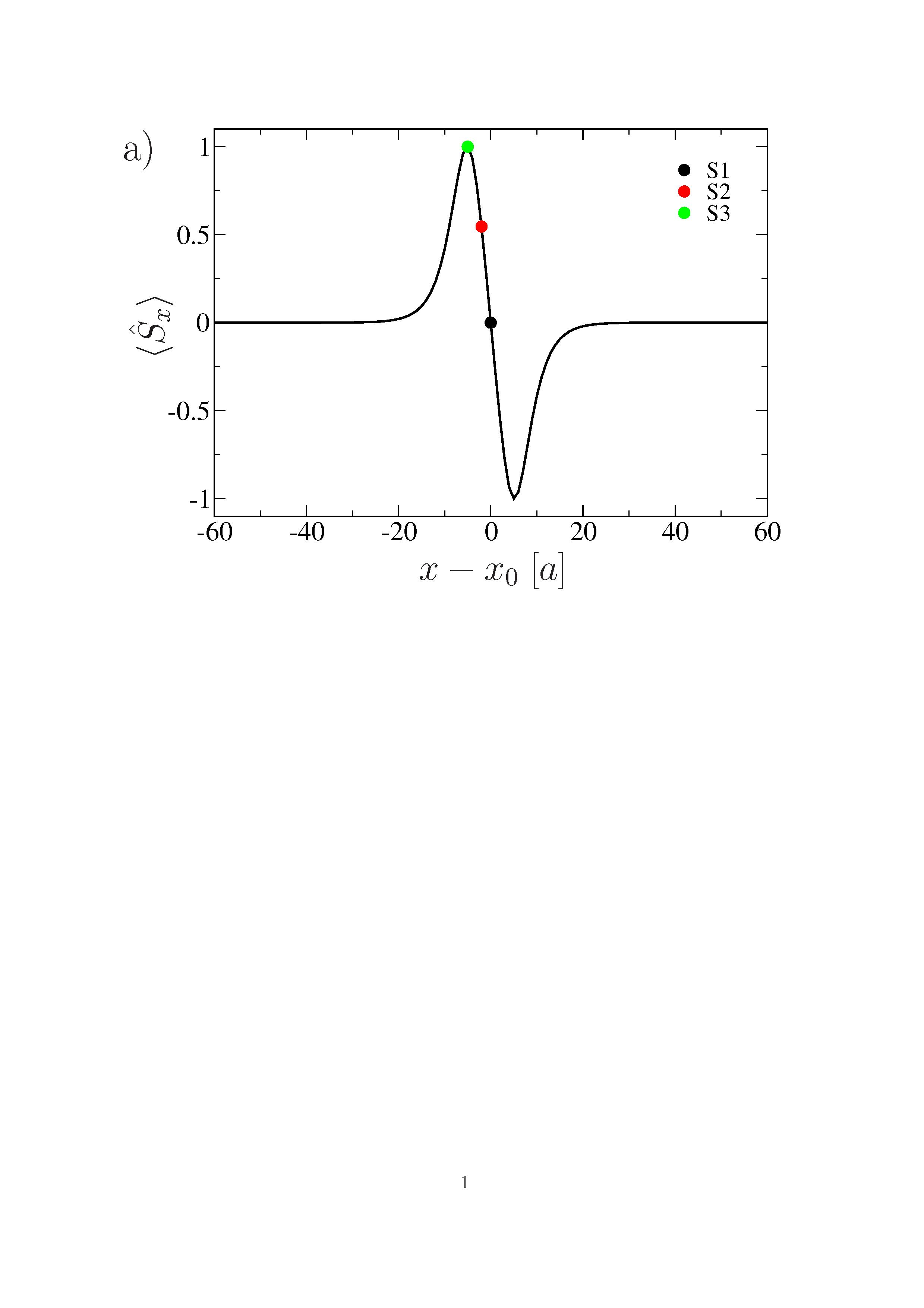} 
\includegraphics*[width=5.75cm,bb = 60 450 520 770]{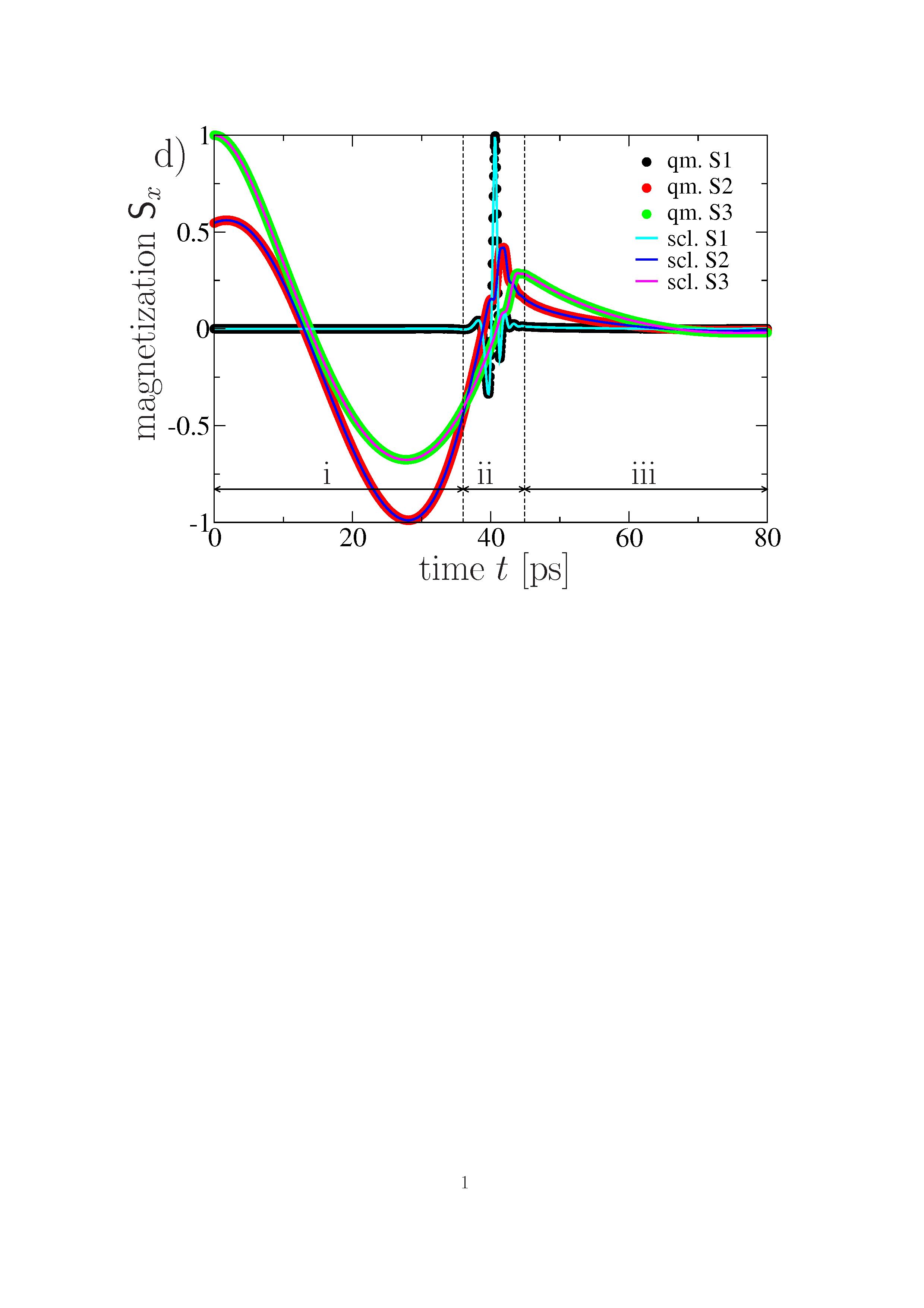} 
\includegraphics*[width=5.75cm,bb = 60 450 520 770]{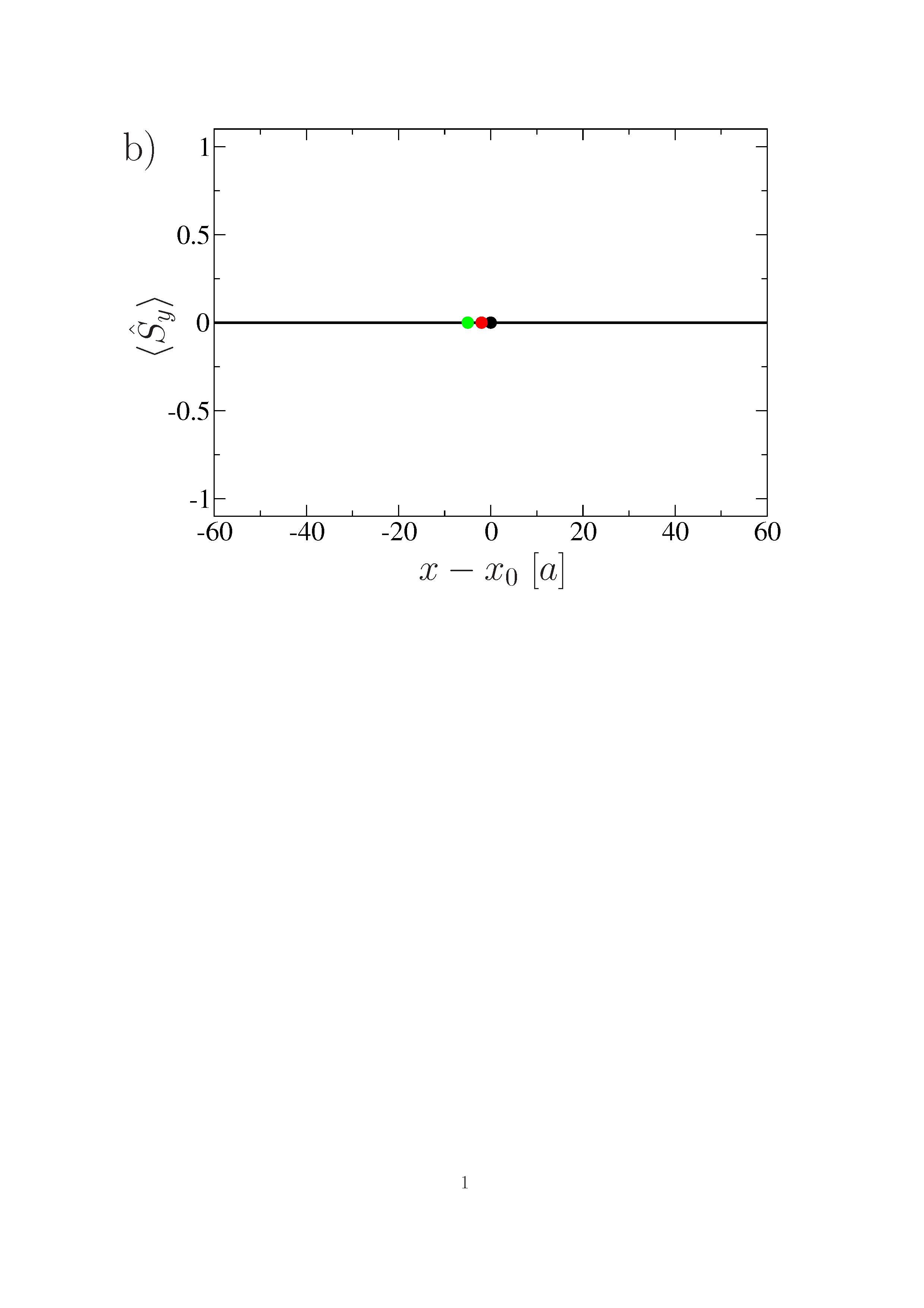}
\includegraphics*[width=5.75cm,bb = 60 450 520 770]{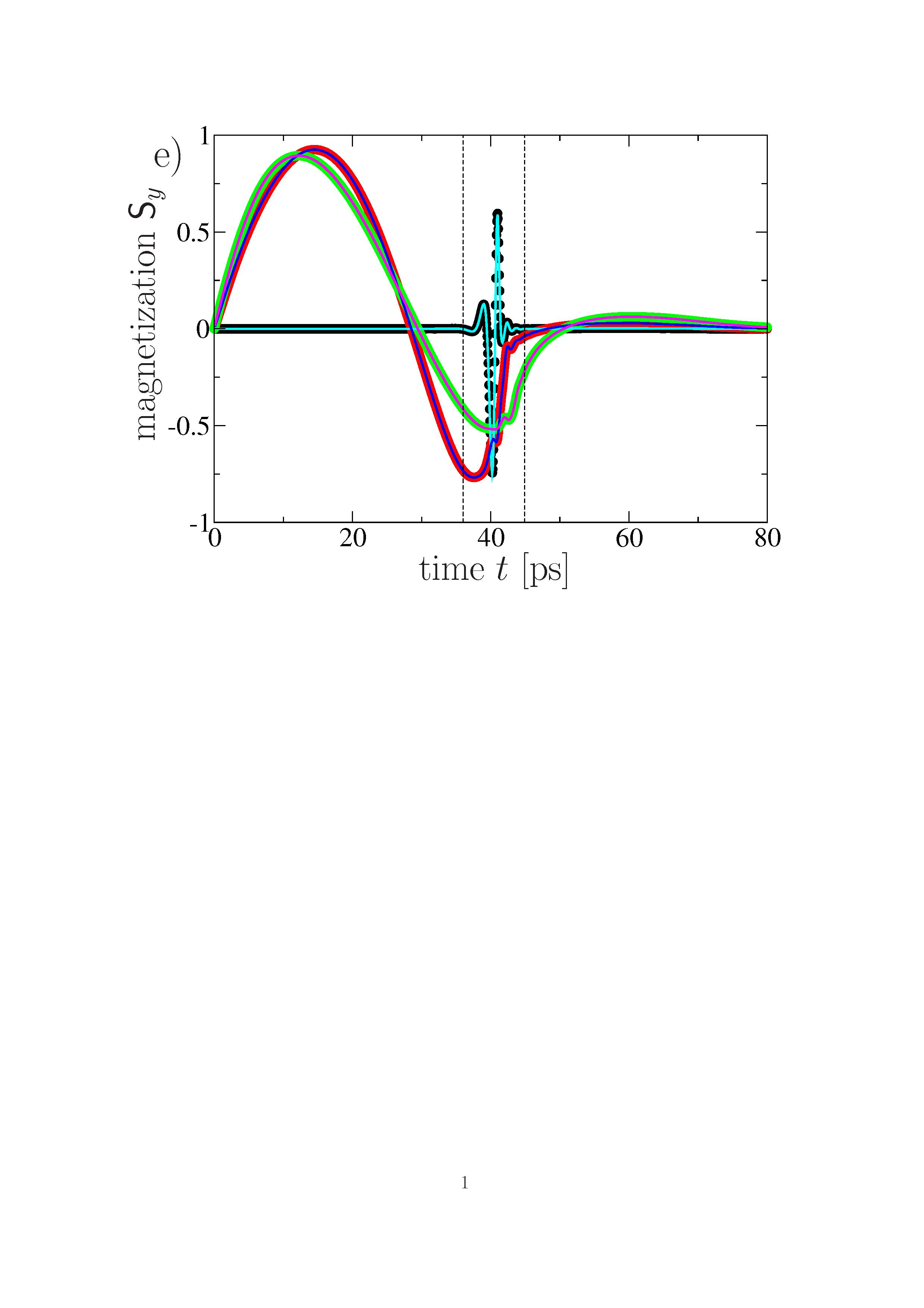} 
\includegraphics*[width=5.75cm,bb = 60 450 520 770]{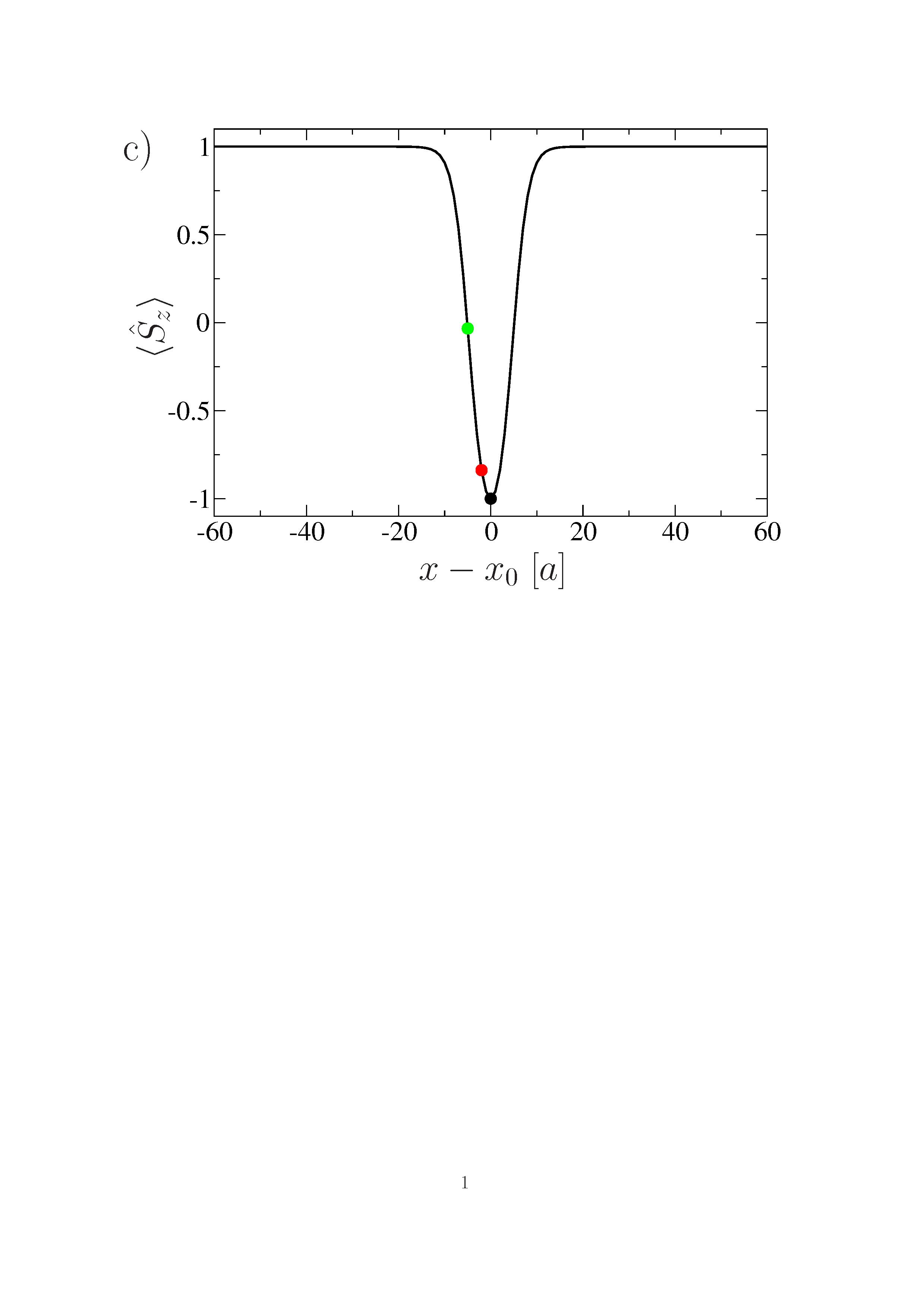} 
\includegraphics*[width=5.75cm,bb = 60 450 520 770]{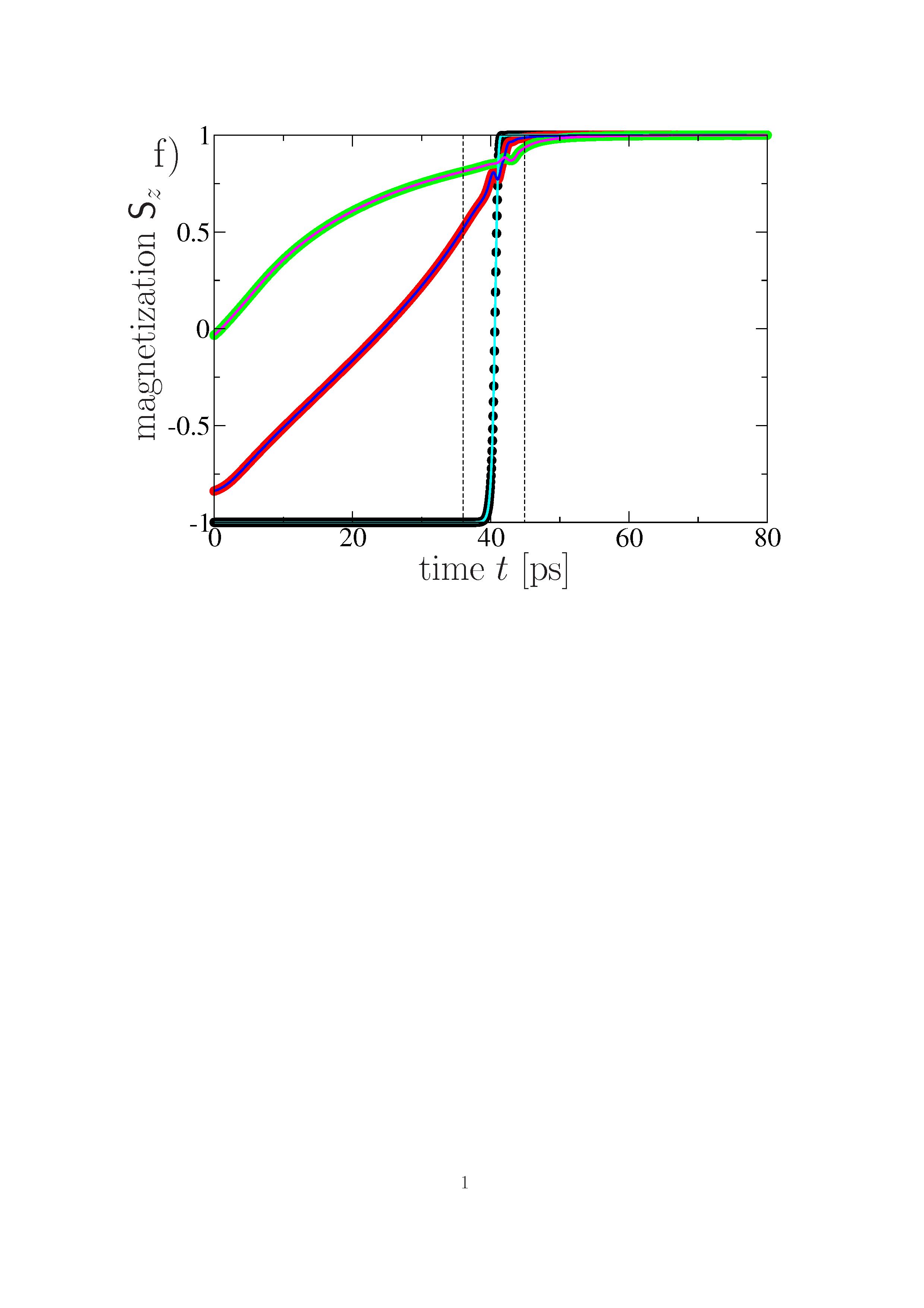} 
  \caption{(color online) a), b), and c) Profiles of the magnetic
    Skyrmion calculated with aid of the time dependent Schr\"odinger
    Eq.~(\ref{tdse2}). d), e), and f) trajectories of three spins
    S1-S3, marked as dots within the Skyrmion profiles, during the
    annihilation process forced by an electric field. The curves are
    calculated by solving the time dependent Schr\"odinger
    Eq.~(\ref{tdse2}) (qm) as well as the semi-classical
    Eq.~(\ref{LLG}) (scl). The dynamics is divided into three 
    phases i: shrinking, ii: collapse, and iii: shock wave.
}       
  \label{f:pic5}
\end{figure*}

\begin{figure*}
\vspace{1mm} % ?????  
\includegraphics*[width=5.cm,bb = 60 450 520 770]{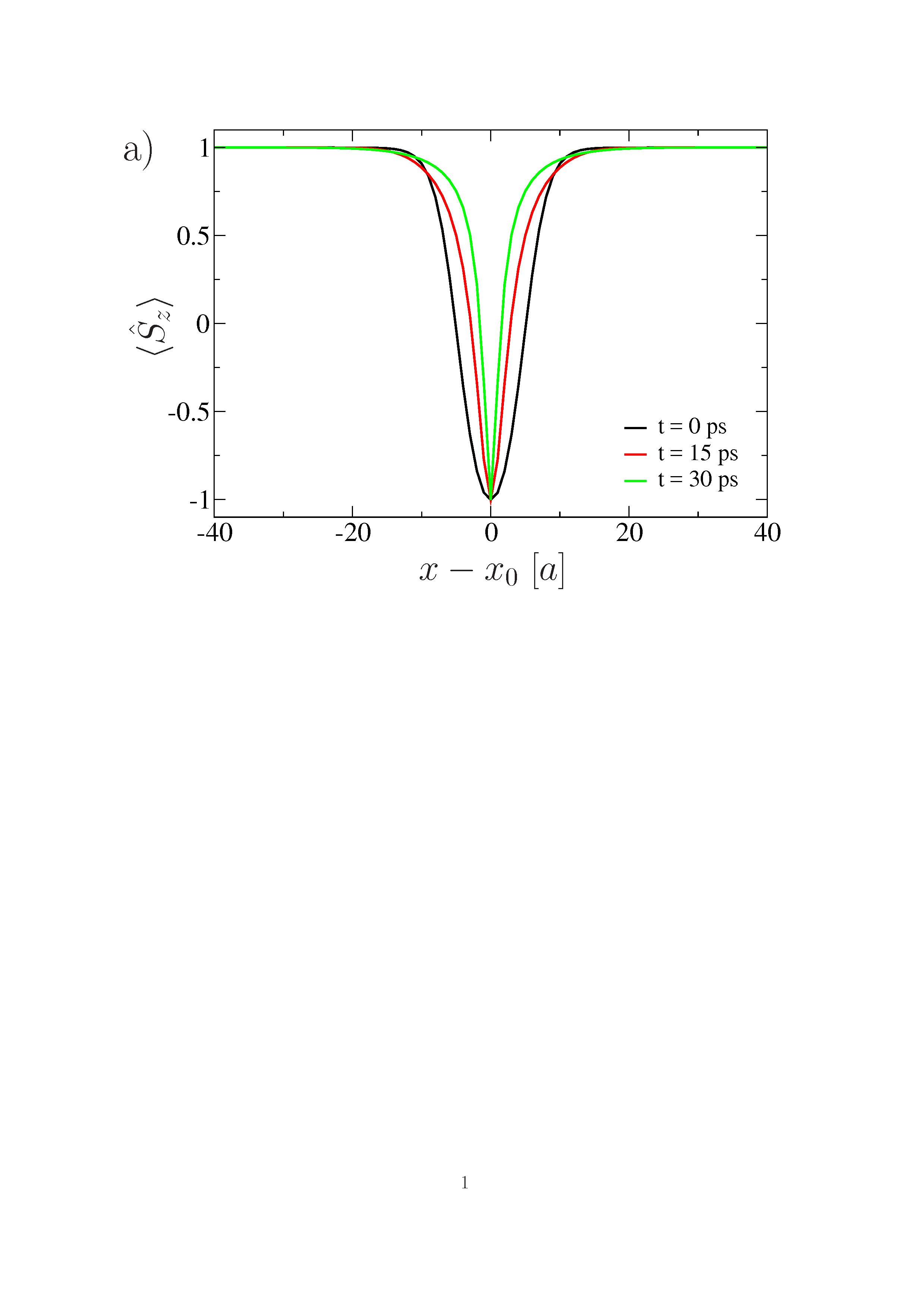} 
\includegraphics*[width=5.cm,bb = 60 450 520 770]{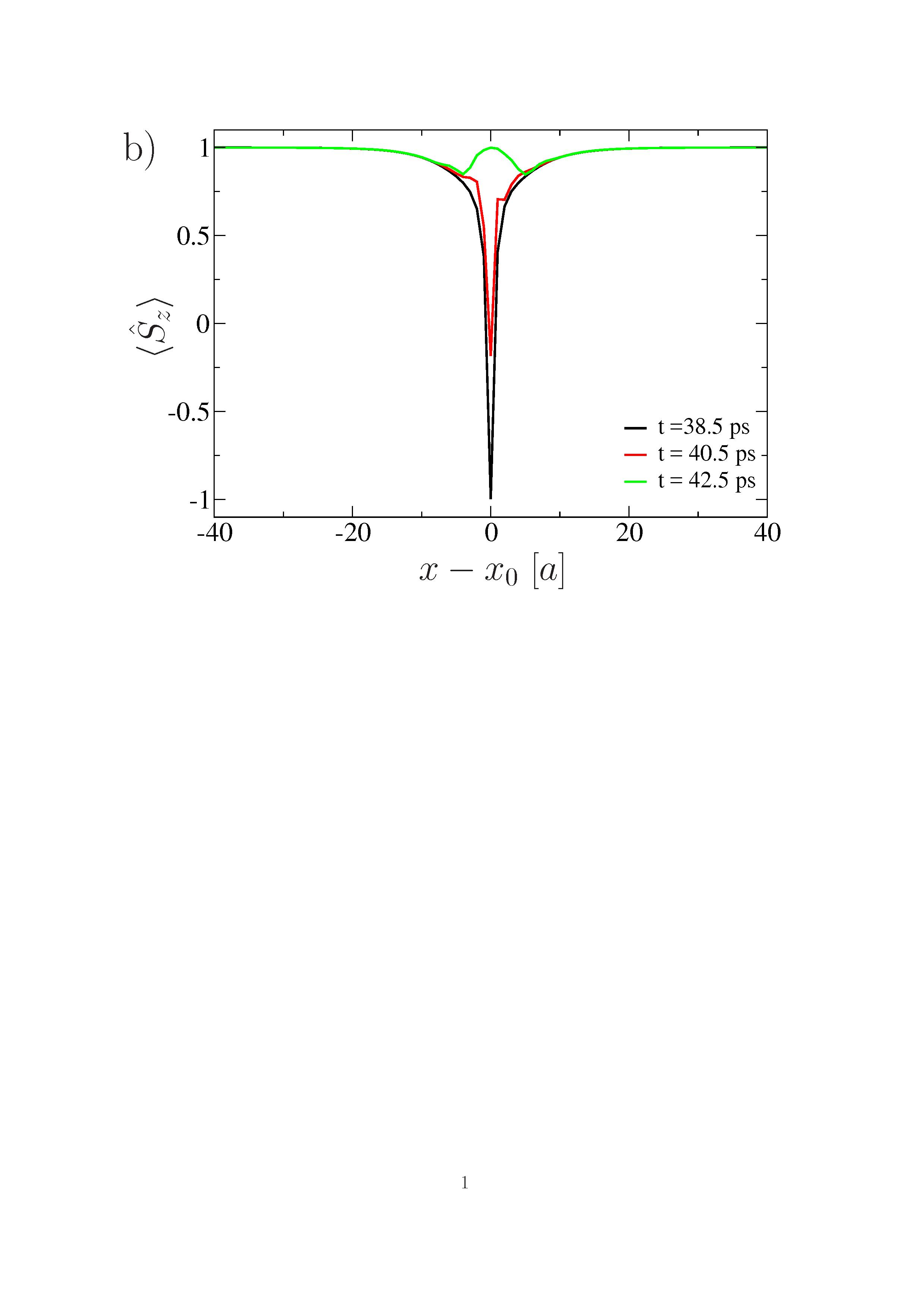} 
\includegraphics*[width=5.cm,bb = 60 450 520 770]{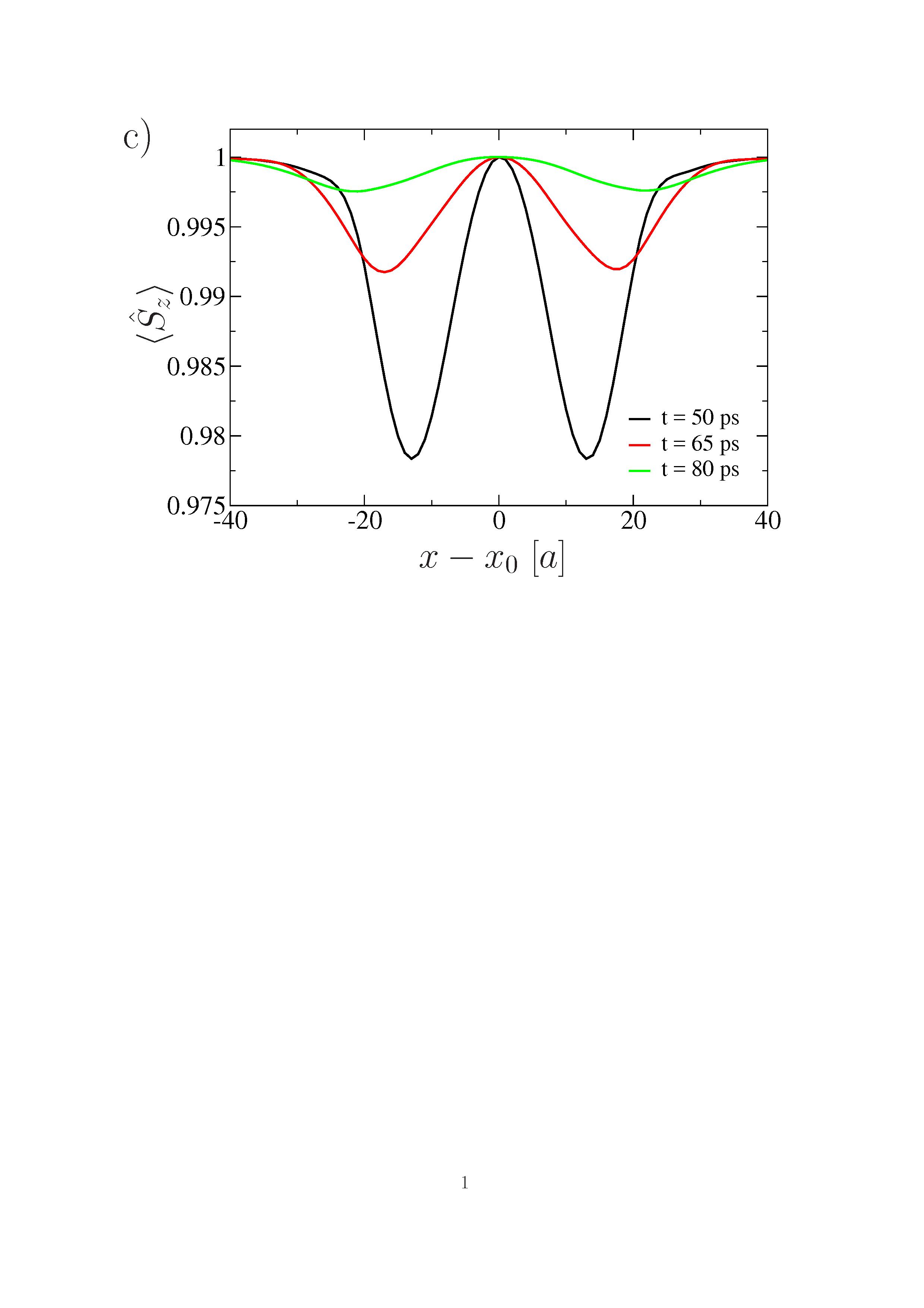}
  \caption{(color online) Skyrmion profiles during the electric field
    forced annihilation process: \\
    a) phase i: shrinking, b) phase ii: collapse, and c) phase iii: shock wave.
}       
  \label{f:pic6}
\end{figure*}

\begin{figure*}
\vspace{1mm} % ?????
\includegraphics*[width=5.cm,bb = 230 565 440 670]{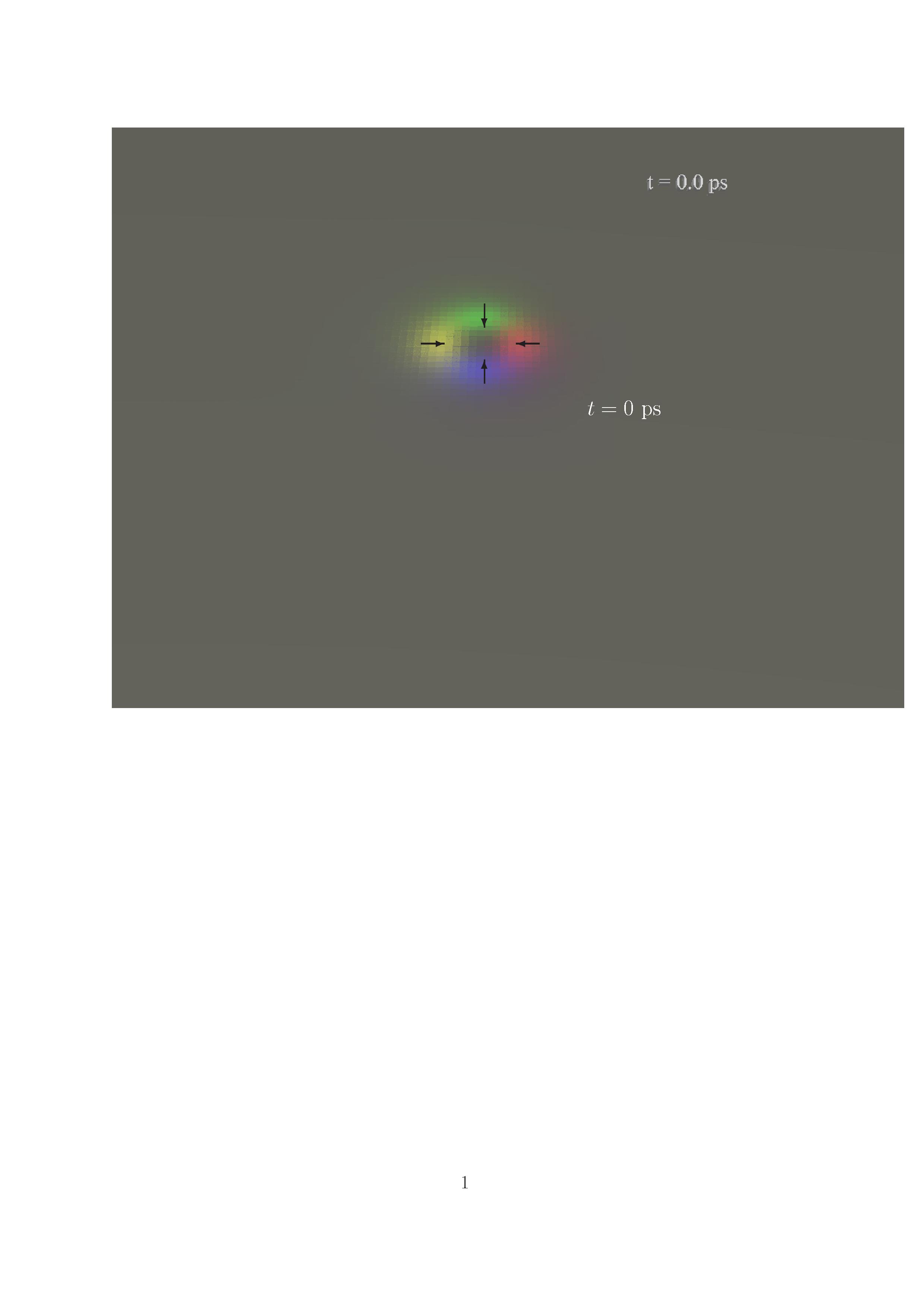}
\includegraphics*[width=5.cm,bb = 230 565 440 670]{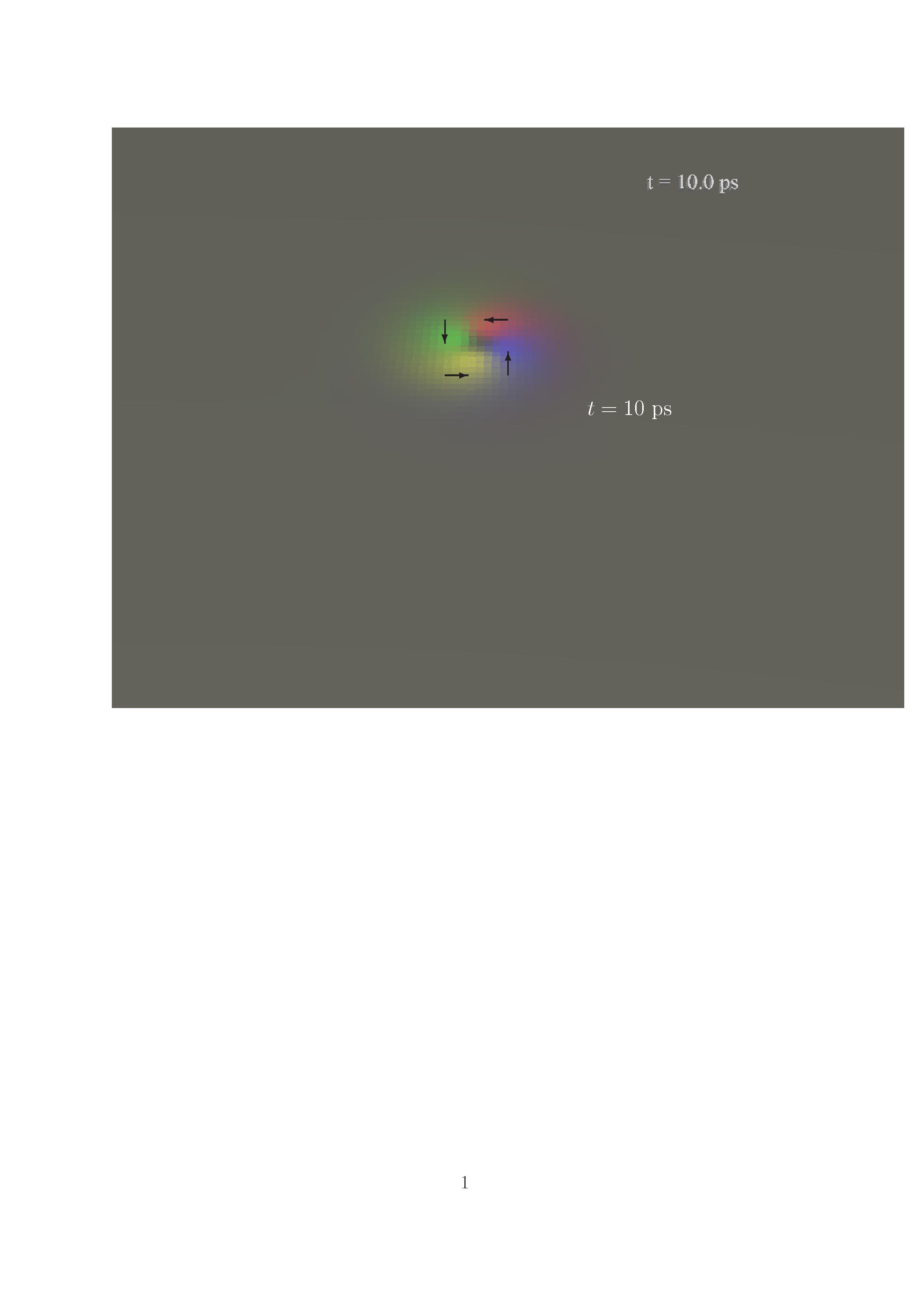}
\includegraphics*[width=5.cm,bb = 230 565 440 670]{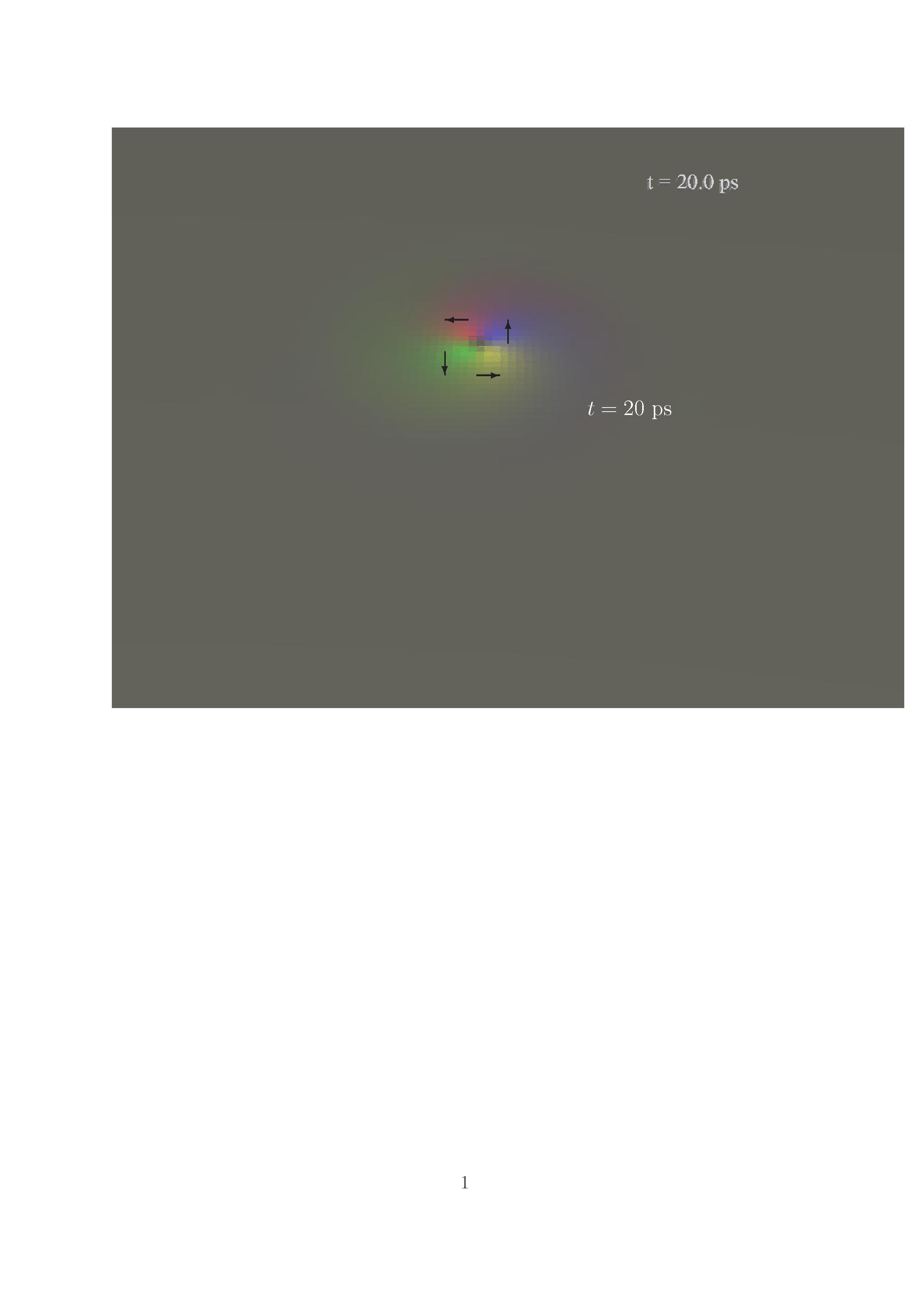}
  \caption{(color online) In-plane components of the magnetic Skyrmion
    during the annihilation process. The small arrows indicate the
    in-plane direction of the magnetization.
}       
  \label{f:pic7}
\end{figure*}

\begin{figure*}
\vspace{1mm} % ?????
\includegraphics*[width=7.5cm,bb = 225 395 600 680]{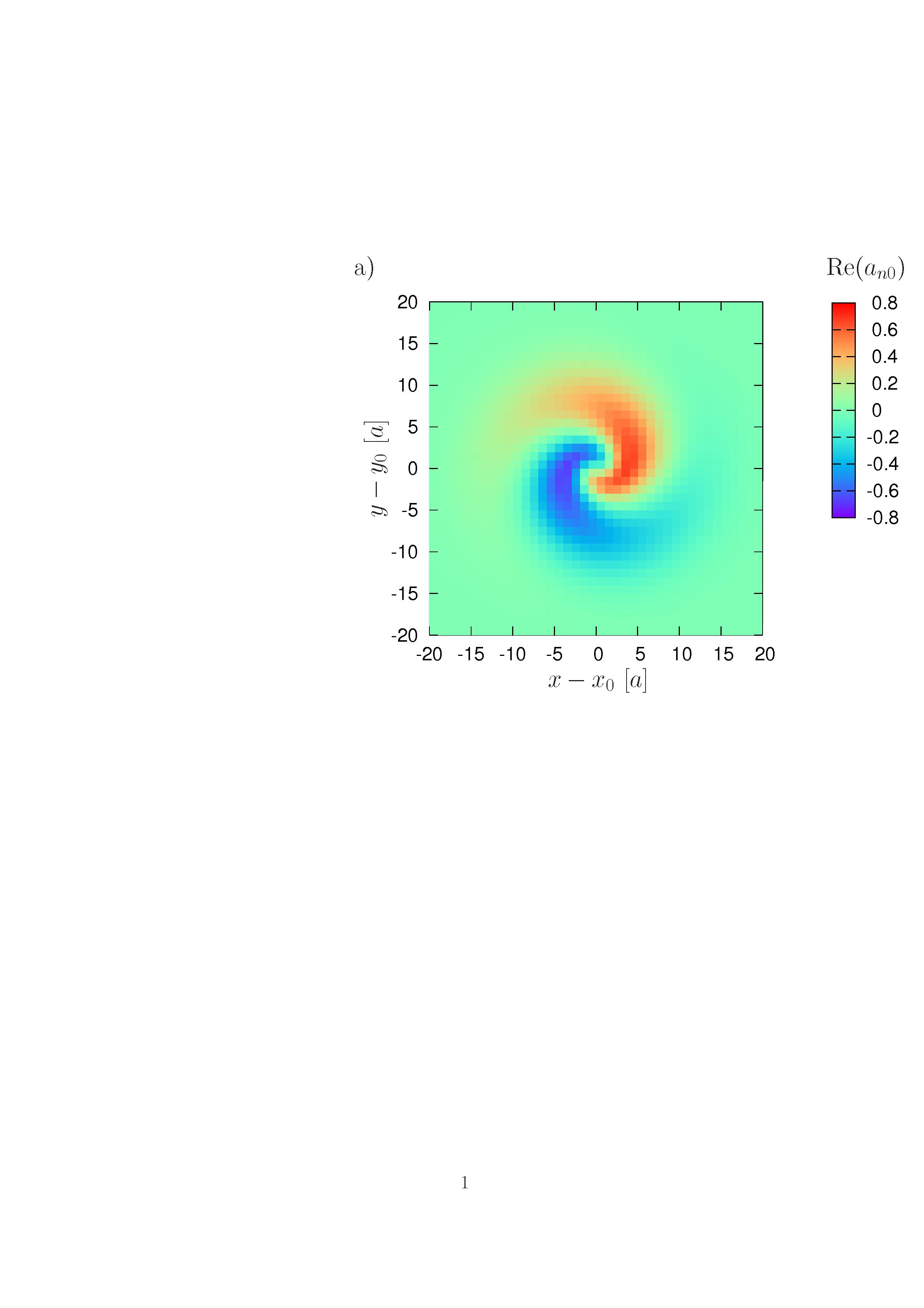}
\hspace{3mm}
\includegraphics*[width=7.5cm,bb = 225 395 600 680]{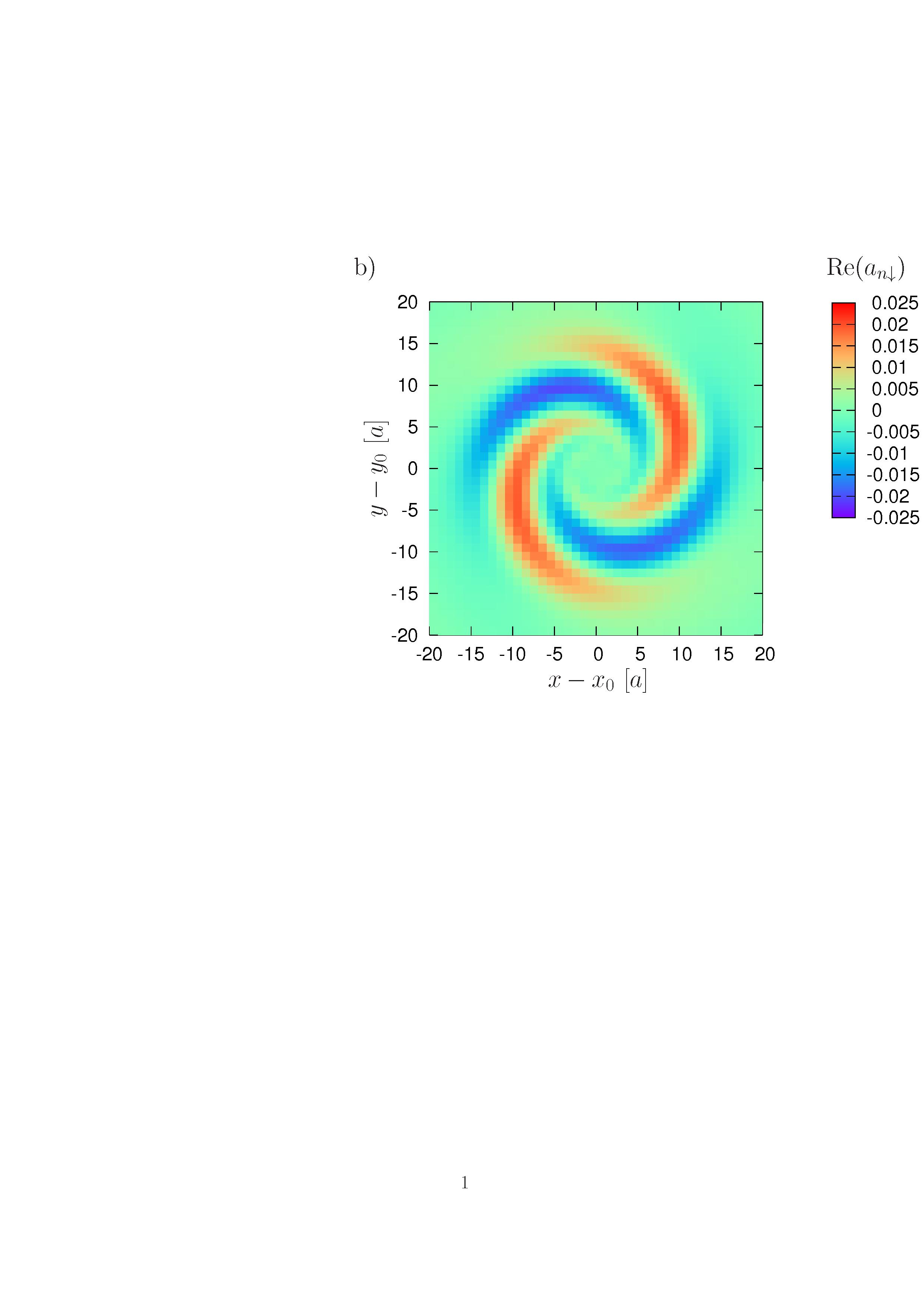}
  \caption{(color online) Real parts of the coefficients $a_{n0}$ and
    $a_{n\downarrow}$ of the wave functions $|\psi_n \rangle$ during
    the electric field forced Skyrmion annihilation. a)
    $\mathrm{Re}(a_{n0})$ at $t = 10$ ps, b)
    $\mathrm{Re}(a_{n\downarrow})$ at $t = 50$ ps.
}       
  \label{f:pic8}
\end{figure*}

Due to the fact that the Hamilton operator
$\hat{\mathrm{H}}$ can be written as sum of the local
Hamilton operators $\hat{\mathrm{h}}_n$ the wave function $|\Psi
\rangle$ is a product state of the local wave functions $|\psi_n\rangle$:
\begin{eqnarray} \label{PSI}
|\Psi \rangle = \bigotimes\limits_{n=1}^N|\psi_n\rangle \;,
\end{eqnarray}
where $N = 200 \times 200 = 40000$. With Eq.~(\ref{HamSCF3}) and
Eq.~(\ref{PSI}) the time dependent Schr\"odinger equation becomes a
set of coupled differential equations:  
\begin{eqnarray} \label{tdse2}
i\hbar (1-\lambda^2) \frac{\mathrm{d}}{\mathrm{d}t}|\psi_n \rangle =
\hat{\mathrm{h}}_n|\psi_n \rangle - i \lambda (
\hat{\mathrm{h}}_n - \langle
\hat{\mathrm{h}}_n \rangle )|\psi_n \rangle \;,
\end{eqnarray} 
where $\langle \hat{\mathrm{h}}_n \rangle = \langle \psi_n |
\hat{\mathrm{h}}_n | \psi_n \rangle$. The wave functions $|\psi_n
\rangle$ can be constructed with aid of the Zeeman basis
\cite{edenCMRA03} where the basis vectors are
$|j\rangle \in \{|\!\uparrow\rangle,|0\rangle,|\!\downarrow\rangle\}$:   
\begin{eqnarray}  
|\psi_n \rangle = \sum_j a_{nj} |j\rangle \;. 
\end{eqnarray}
In case the system is in the ground state the coefficients $a_{nj}$
are equal to $\phi^-_{nj}$ and the wave functions $| \psi_n \rangle$
are equal to the eigenvectors ${\boldsymbol \phi}_n^-$ (see
Sec.~\ref{s:EF}). Due the fact that the wave function $|\Psi \rangle$
is a product state [see Eq.~(\ref{PSI})] the system shows no
entanglement. Therefore we can expect a spin dynamics which is similar
to the dynamics provided by a description using classical spins
\cite{wieserJPCM16}.   

In general the interest in the wave functions $|\psi_n \rangle$ is
restricted despite the fact that the wave functions can provide 
additional information about the quantum system. More important and
necessary for the dynamics are the spin expectation values:   
\begin{eqnarray} 
\langle \hat{\mathbf{S}}_n \rangle = \langle \Psi
|\hat{\mathbf{S}}_n |\Psi \rangle = \langle \psi_n
|\hat{\mathbf{S}}_n |\psi_n \rangle \;. 
\end{eqnarray}
In the previous publications \cite{wieserEPJB15,wieserJPCM16} it
has been shown that without entanglement the dynamics of the spin
expectation values are well described by the following differential equation: 
\begin{eqnarray} \label{LLG}
(1-\lambda^2) \frac{\mathrm{d}}{\mathrm{d}t} \langle
  \hat{\mathbf{S}}_n \rangle = \gamma \langle \hat{\mathbf{S}}_n 
  \rangle \times (\mathbf{B}_n^{\mathrm{eff}} - \lambda
  (\langle \hat{\mathbf{S}}_n \rangle \times
  \mathbf{B}_n^{\mathrm{eff}}) ) \;, \nonumber \\
\end{eqnarray}
where $\gamma = g\mu_B/\hbar$ is the gyromagnetic ratio. The important
point here is that due to the absence of entanglement the absolute
values $|\langle \hat{\mathbf{S}}_n \rangle| = \hbar S$ are time
independent (constant). The spin expectation values $\langle
\hat{\mathbf{S}}_n \rangle$ only change their orientations in spin
space. This is similar to the dynamics of the classical spins
$\mathbf{S}_n$ described by the Landau-Lifshitz-Gilbert (LLG)
equation. Indeed, Eq.~(\ref{LLG}) is identical to the LLG equation
just with a tiny difference: the sense of rotation 
of the precession is reversed (for details see \cite{wieserJPCM16}). 

Fig.~\ref{f:pic5} shows the trajectories of three spins during the
annihilation of a Skyrmion due to the external electric
field. Hsu et al. \cite{hsuARXIV16} have demonstrated that it
is possible to use a local electric field generated by a scanning
tunneling microscope to create or annihilate magnetic
Skyrmions. However, the theoretical explanation for this phenomenon is
is missing within the publication of Hsu et al. which is the tuning of the
Dzyaloshinsky Moriya interaction by the electric field
\cite{wieserARXIV16}: 
\begin{equation}
{\boldsymbol{\cal D}}_{nm} = {\boldsymbol{\cal D}}_{nm}^0 +
    {\boldsymbol \omega}_{nm} (\mathbf{E} \times \mathbf{r}_{nm}) \;.
\end{equation}
Within this formula the ${\boldsymbol{\cal D}}_{nm}$ are the Dzyaloshinsky
Moriya vectors after modification due to the electric field
$\mathbf{E}$. The original Dzyaloshinsky Moriya vectors are
${\boldsymbol{\cal D}}_{nm}^0$ while $\mathbf{r}_{nm}$ is the vector
pointing from lattice site $m$ to lattice site $n$. The prefactor
${\boldsymbol \omega}_{nm}$ is a constant which describes the strength of the
modification due to the electric field. In the case of the experiment
using the scanning tunneling microscope the modification of the
Dzyaloshinsky Moriya vectors is local underneath the scanning
tunneling microscope tip and the electric field vector $\mathbf{E}$
perpendicular to $\mathbf{r}_{nm}$. Depending on the orientation and
strength of $\mathbf{E}$ the Dzyaloshinsky Moriya interaction becomes
increased or reduced. Within the spin dynamics simulations presented
in the following a constant modification ${\boldsymbol \omega}_{nm} =$
const. within a radius of $20$ lattice sites is assumed in such a way
that outside this circle the strength of the Dzyaloshinsky Moriya
interaction is not modified ${\boldsymbol \omega}_{nm} = 0$ while
inside the circle the Dzyaloshinsky Moriya interaction is zero
${\boldsymbol{\cal D}}_{nm} = 0$. This modification simulates the
influence of the electric field provided by the scanning tunneling
microscope tip in the experiment by Hsu et al.. Thereby, the field
vector $\mathbf{E}$ is perpendicular to the film plane $\mathbf{E}
\perp \mathbf{r}_{nm}$ pointing to the film plane. Without
Dzyaloshinsky Moriya interaction the ferromagnetic state 
is the ground state and the Skyrmion gets annihilated. Additional,
within the spin dynamics simulation a small misalignment of the center
of the Skyrmion and the magnetic tip (center of the circle) $0.5$ lattice
constants in both directions $x$ and $y$ has been assumed. The small
misalignment has the reason to break the symmetry of the central spin
which otherwise would have no distinguished spin torque during the
annihilation process. During the experiment by Hsu et al. the symmetry
is broken either by thermal fluctuations or by also by
misalignment respectively assymmetry of the electric field due to an
non-symmetric tip. Fig.~\ref{f:pic5} provides the trajectories of the
three spins mentioned before. The spins are marked  
by the colored dots in Fig.~\ref{f:pic5}a)-c), where
Fig.~\ref{f:pic5}a) provides the Skyrmion profile corresponding to the
$x$-component, Fig.~\ref{f:pic5}b) the $y$-component, and
Fig.~\ref{f:pic5}c) the $z$-component of the spin expectation value $\langle
\hat{\mathbf{S}}_n \rangle$. The trajectories in
Fig.~\ref{f:pic5}d)-f) are calculated in two ways: 1. solving the time 
dependent Schr\"odinger Eq.~(\ref{tdse2}) and 2. solving the
semi-classical differential Eq.~(\ref{LLG}). The remarkable fact is
that the trajectories of both calculations show a perfect
agreement. At this point it is necessary to mention that the 
agreement is a consequence of the absent entanglement. 

Now the question is what are the advantages and disadvantages of using
the semi-classical differential Eq.~(\ref{LLG}) instead of solving the
time dependent Schr\"odinger equation? The advantage of the semi-classical
differential equation is the fact that we have to calculate only three
(one for $\langle \hat{S}_n^x \rangle$, $\langle \hat{S}_n^y \rangle$,
and $\langle \hat{S}_n^z \rangle$) instead of six (real and imaginary part of
$a_{n\uparrow},a_{n0},a_{n\downarrow}$) differential equations per
lattice site. Furthermore, the spin expectation values $\langle
\hat{\mathbf{S}}_n \rangle$ are directly given while for the time
dependent Schr\"odinger equation first the wave functions
$|\psi_n\rangle$ have to be calculated and in a second step the
expectation values $\langle \hat{S}_n^\eta \rangle$, $\eta \in
\{x,y,z\}$. Therefore, it can be said that in summary the effort and
also the possible numerical errors are reduced by using the
semi-classical differential Eq.~(\ref{LLG}) instead of the time 
dependent Schr\"odinger Eq.~(\ref{tdse2}). However, on the other hand,
Eq.~(\ref{LLG}) does not provide the wave functions $|\psi_n
\rangle$. This means that the information, delivered by 
the wave functions, get lost. Furthermore, in the case of entanglement it is
necessary to solve the time dependent Schr\"odinger equation. 
%At least not directly%

So far we got a rough idea about the annihilation process provided by
Fig.~\ref{f:pic5}. The process itself can be divided into three
phases marked by the roman numbers: The first phase (phase i) is
characterized by the shrinking of the Skyrmion. During the
second phase (phase ii) the Skyrmion collapses. And the third phase
(phase iii) is a shock wave running trough the system. Corresponding to
this dynamics Fig.~\ref{f:pic6} shows the profile of the Skyrmion
during these three phases: Fig.~\ref{f:pic6}a) corresponds to phase i,
Fig.~\ref{f:pic6}b) to phase ii and Fig.~\ref{f:pic6}c) to phase
iii. Fig.~\ref{f:pic7} shows the in-plane components of the
magnetization. Clearly visible the twist of the Skyrmion during the
annihilation process. This twist can be seen also in the wave
functions $|\psi_n \rangle$. Fig.~\ref{f:pic8} shows the real parts of
$a_{n0}$ and $a_{n\downarrow}$ during the first (i) and third (iii)
phase. 

\section{Summary} \label{s:summary}

The paper is separated in three parts where a quantum mechanical SCF
theory has been used to describe the thermodynamics, the ground state 
wave function and spin dynamics of a magnetic Skyrmion. The first part
describes the thermodynamics. The main results of this part are the
temperature dependent profile of the Skyrmion and the phase
diagram. The later is a result of the analysis of first 
order transition between the Skyrmionic and ferromagnetic phase as
well as the second order phase transition between the ferromagnetic
respectively spin-spiral phase and the paramagnetic phase. The second
part provides the eigenvalues of the Hamiltonian
$\hat{\mathrm{h}}_n$ which can be used as the starting 
point for the spin dynamics. The third part describes the quantum spin
dynamics of the Skyrmion. It is shown that the trajectories of the
spin expectation values $\langle \hat{\mathbf{S}}_n \rangle$ can be
described either by the time dependent Schr\"odinger Eq.~(\ref{tdse1})
or the semi-classical differential Eq.~(\ref{LLG}). The description
using the time-dependent Schr\"odinger equation is similar to the one
of the time-dependent Hartree method \cite{grossmannBOOK08} and the
advantage of using the semi-classical differential Eq.~(\ref{LLG}) is
the reduced numerical effort. On the other hand the time dependent
Schr\"odinger equation provides additional informations via the wave
functions. Concerning the physics: The investigated 
scenario is the annihilation process of the Skyrmion using an external
electric field. The process of annihilation itself can be separated into
three phases: size reduction of the Skyrmion, collapse and resulting
shock wave due to the energy gain (first order phase transition).

%\begin{acknowledgments}
%Nada 
%\end{acknowledgments}

\bibliography{Cite}

\end{document}